\documentclass[12pt]{article}
\usepackage{setspace}
\usepackage{amsmath}
\usepackage{amssymb}
\usepackage{graphics,psfrag,epsfig,rotating}
\usepackage{subfigure}
\usepackage{overcite}

\newcommand{\tl}{\tilde}

\begin{document}

\title{
\vspace{-15mm}
Investigations of the NS-$\alpha$ model using a lid-driven cavity flow}

\author{K. A.~Scott\\
   Department of Mechanical Engineering, University of Waterloo,\\
   Waterloo, Ontario N2L 3G1, Canada\\
   E-mail: ka3scott@engmail.uwaterloo.ca
 \and   F. S.~Lien\\
   Department of Mechanical Engineering, University of Waterloo,\\
   Waterloo, Ontario N2L 3G1, Canada}
\date{}

\maketitle

\maketitle

\begin{abstract}

In this paper we investigate a subgrid model based on an anisotropic version of the NS-$\alpha$ model using a lid-driven cavity flow at a Reynolds number of 10,000.  Previously the NS-$\alpha$ model has only been used numerically in the isotropic form. The subgrid model is developed from the Eulerian-averaged anisotropic equations [Holm, \textit{Physica D}, v.133, pp 215-269, 1999]. It was found that when $\alpha^{2}$ was based on the mesh numerical oscillations developed which manifested themselves in the appearance of streamwise vortices and a `mixing out' of the velocity profile. This is analogous to the Craik-Leibovich mechanism, with the difference being that the oscillations here are not physical but numerical. The problem could be traced back to the discontinuity in $\alpha^{2}$ encountered when $\alpha^{2}=0$ on the endwalls. An alternative definition of $\alpha^{2}$ based on velocity gradients, rather than mesh spacing, is proposed and tested. Using this definition the results with the model shown a significant improvement. The splitting of the downstream wall jet, rms and shear stress profiles are correctly captured a coarse mesh. The model is shown to predict both positive and negative energy transfer in the jet impingement region, in qualitative agreement with DNS results. 
\end{abstract}

\section{Introduction}
\label{intro}
An accurate description of turbulent flows is of paramount importance both in terms of engineering applications, and in understanding physical phenomena in the natural world. Increasingly, numerical computations are playing a prominent role in turbulence research. However, for many practical problems, the full range of scales active in a turbulent flow cannot be resolved on a finite computational domain. This means models must be introduced to parameterize the effects of the unresolved motions on the resolved ones. This is usually done by first applying averaging procedures directly to the Navier-Stokes equations. The most common methods are either to introduce a statistical average, which leads to the Reynolds-Averaged Navier-Stokes (RANS) equations, or to use a spatial filter, which leads to the Large-Eddy Simulation (LES) equations. Both methods lead to the appearance of an unclosed term, for which a wide variety of models have been presented for both RANS and LES methodologies\cite{Hanjalic2005,Sagaut2002}. One consistent trend has been the use of models which are either strictly dissipative, or contain a dissipative component. This concept is well-founded, since the role of the small scales, which are being modeled, is to remove the energy generated through non-linear interactions of the large, resolved scales. However, there are some flows where the non-linear interactions are weak, and the dissipation provided by such models may be excessive. An example of this is the early stages of transition in a boundary layer flow, where dissipative models may delay, or even prevent, the onset of transition \cite{Piomelli1991}. 
\\
\\
The NS-$\alpha$ model is considered a non-dissiptive model in that it uses a modified nonlinearity to alter the energy transfer among scales instead of changing the dissipative process \cite{Graham2008}. However, it has different origins than other nonlinear models. Instead of starting with the Navier-Stokes equations, the governing equations can be derived by applying Hamilton's principle to an averaged Lagrangian \cite{Holm1999}, and the resulting (inviscid) equations conserve circulation when the model parameter $\alpha^{2}$ is constant. The derivation using Hamilton's principle naturally leads to a set of equations which contain two velocity fields $u_{i}$ and $\tl{u}_{i}$, where $\tl{u}_{i}$ is smoother than $u_{i}$ via an inversion of the Helmholtz operator. In the isotropic case the the parameter that arises in the averaging procedure is a scalar, $\alpha^{2}$.  When $\alpha^{2}$ is constant the governing equations can be written \cite{Chen1999c}
\begin{equation}
\partial_{i} \tl{u}_{i}=0,
\end{equation}
\begin{equation}
\partial_{t} u_{i} + \tl{u}_{j}\partial_{j} u_{i} + u_{k}\partial_{i} \tl{u}_{k}= -\partial_{i} p^{\alpha}+\nu\partial_{kk} u_{i},
\label{NSalpha_iso}
\end{equation}
with,
\begin{equation}
u_{i}=\left(1-\alpha^{2}\partial_{kk}\right)\tl{u}_{i},
\end{equation}
\begin{equation}
 p^{\alpha}=p-\frac{1}{2}u_{i}\tl{u}_{i}.
\end{equation}
The third term on the LHS of \eqref{NSalpha_iso} is unique to the NS-$\alpha$ model and will be referred to in the following as the tilting term because it arises due to the velocity difference between two ends of a Lagrangian trajectory which is being carried by a smoothed flow. We also use the standard Laplacian operator acting on the momentum velocity $u_{i}$ in the dissipation term in the interest of maintaining a model similar to that used in other studies \cite{Geurts2006,Holm2003,Chen1999c,Graham2008}.  The above set of equations are also known as the viscous Camassa-Holm equations. Interest in using these equations as a model for turbulence can be traced back to the work by Chen \cite{Chen1998,Chen1999c} where analytical results shown to yield velocity and shear stress profiles in good agreement with experimental results for pipe and channel flows. 
\\
\\
The early literature on the NS-$\alpha$ equations hypothesized that the equations would have an energy spectrum with a steeper slope in the inertial subrange for length scales smaller than $\alpha$ \cite{Chen1999a,Chen1999b,Foias2001}, making it a good candidate for an LES model. The slope based on the conserved energy $E_{\alpha}=\int_{V} \tl{u}_{i}u_{i} \, dV$ was expected to be $k^{-1}$, which corresponds to $k^{-3}$ for the translational energy $E_{\tl{u}}=\int_{V} \tl{u}_{i}\tl{u}_{i} \, dV$. The physical mechanism behind the steeper slope was explained as  the suppression of nonlinear interactions between scales which are smaller than $\alpha$ \cite{Domaradzki2001}.  More recently it has been shown in an enlightening study \cite{Graham2008} that the energy spectrum is in fact not steeper than that of the Navier-Stokes equations. The reason for this is that the NS-$\alpha$ fluid is comprised of both regions undergoing the Navier-Stokes dynamics of vorticity transport and stretching, and regions described as `rigid rotators' \cite{Graham2008} where stretching is inhibited. These rigid rotators have no internal degrees of freedom but do have kinetic energy. Their scaling leads to an energy spectrum which has a slope of $k^{1}$, which means the $k^{-1}$ spectrum is subdominant \cite{Graham2008}.  Thus while the hypothesis of reduced nonlinear activity at the small scales appears correct, the scaling of the observed spectrum is different than what was anticipated.
\\
\\
A similar principle to reduce small-scale activity is used in the Leray model\cite{Geurts2003}, which is based on Leray's regularization of the Navier-Stokes equations\cite{Leray1934}. For the Leray model the momentum equations are the same as the Navier-Stokes equations with the exception that the advecting velocity is smoothed. Thus, they have the same form as equation \eqref{NSalpha_iso} with the third term on the LHS set to zero and $p^{\alpha}=p$. 
However, in the Leray model it is the incompressibility of the unfiltered velocity which is enforced, $\partial_{i}u_{i}$. It has been shown that this leads to significant problems when the model is used for wall-bounded flows \cite{vanReeuwijk2006}.  
\\
\\
The reduced small-scale activity of the NS-$\alpha$ and Leray equations have led to the suggestion \cite{Geurts2003} to use these equations as turbulence models. This can be done either by working directly with the equations in which the unsmoothed velocity is the dependent variable, for example \eqref{NSalpha_iso}, or by rewriting these equations in terms of the smoothed velocity $\tl{u}_{i}$. The latter is considered an LES methodology and gives rise to a governing equation that has the standard LES template \cite{Geurts2003}. This is the approach that will be investigated here. Previous studies along these lines include the temporal transition of a mixing layer \cite{Geurts2006}, the parameterization of mixing in a gyre \cite{Holm2003}, and studies of decaying and forced box turbulence \cite{Mohseni2003}. A conclusion that can be drawn from these studies is that the NS-$\alpha$ model captures the variability of the large, resolved scales better, \cite{Geurts2006} or at least as well as\cite{Mohseni2003}, the dynamic model. However, if the grid resolution is too coarse there will be a build-up of energy associated with the subgrid fluctuations. This is particularly severe if the initial condition contains a broadband spectrum \cite{Mohseni2003}. These results are not surprising in light of the fact that while the NS-$\alpha$ model attenuates triad interactions associated with the forward energy transfer, it does not do so abruptly at a wavenumber corresponding to $k_{\alpha}=1/\alpha$. It is expected that a scale separation between the grid scale cut-off and $\alpha$ would be required to allow for this attenuation. If we consider that we still need to resolve the scales that eddies of scale $1/\alpha$ transfer their energy to, this means $k_{max}=2/\alpha$. Since the maximum wavenumber is also related to the grid spacing as $k_{max}\sim \pi/h$ this gives, $\alpha/h \sim 2/\pi \sim 1$. This is in agreement with the subgrid resolution suggested by Geurts and Holm \cite{Geurts2006}, which they determined through grid-refinement studies.
\\
\\
While it is possible to define guidelines for how big $\alpha^{2}$ should be in the case of homogeneous, isotropic turbulence, it is not clear how to proceed in the more general situation. Since $\alpha^{2}$ is the only parameter in this model we expect its specification to be critical. Physically $\alpha^{2}$ can be interpreted either in the Lagrangian sense as a measure of rms particle displacement, or in the Eulerian sense as a mixing length. Numerically, it can be interpreted as a filter width \cite{Geurts2006}. The studies noted above all used a constant value for $\alpha^{2}$, which was taken to be a fraction of the domain size. It can be expected that there will be many situations where it may not be appropriate to maintain a constant value of $\alpha^{2}$, and in addition, where we may want $\alpha^{2}$ to reflect the anisotropy of the flow. For example, near a solid wall we would expect the particle displacement and mixing length normal to the wall to decrease. In a similar manner, in LES we usually reduce the filter width and refine the grid in this region. 
\\
\\
As noted by Zhao and Mohseni \cite{Zhao2005a} there are two possible ways to proceed with the problem of specifying $\alpha^{2}$ in an anisotropic flow. One way would be to use the anisotropic NS-$\alpha$ equations \cite{Holm1999,Marsden2003} to develop an equation for $\tl{u}_{i}$. A second approach would be to use the isotropic equations, but modify $\alpha^{2}$ to reflect the anisotropy of the flow. Zhao and Mohseni followed the second approach, and formulated a dynamic procedure to specify $\alpha^{2}$. The model was tested $\textit{a priori}$ on channel flow, where $\alpha^{2}$ was found to be constant away from the wall, decaying to zero such that $\alpha^{2}=0$ at the wall. However, \textit{a posteriori} results were less encouraging \cite{Zhao2005b}. 
\\
\\
Outside of these studies by Zhao and Mohseni using the isotropic model, the author is not aware of any results in the literature where the NS-$\alpha$ model has been used in a numerical simulation of wall-bounded flows of interest to engineers (although the Leray-$\alpha$ model has been studied \cite{vanReeuwijk2007}, as have related symmetry-preserving methods \cite{Verstappen2007}). Given the importance of this problem, and the promise the model is showing for unbounded flow, the objective of this study is to contribute to this picture by investigating using anisotropic the NS-$\alpha$ equations as a subgrid model.  The anisotropic equations are comprised of a set of coupled PDEs governing momentum conservation and the particle displacement covariance \cite{Holm1999,Marsden2003}. For the Eulerian-averaged equations investigated here, the governing equation for the displacement covariance is simply an advection equation, and it is not clear how such an equation would be initialized, or even if it should be treated in the prognostic sense.  Instead of solving this equation, the initial approach taken here was to view $\alpha^{2}$ as a smoothing scale which is based on the grid, and assess the performance of the model based on this definition. Unfortunately, difficulties were encountered with using a simple mesh-based specification. For this reason an alternative definition of $\alpha^{2}_{k}$ was tested. It should be noted that results using the isotropic NS-$\alpha$ model are not presented here. We found this model generates excessive backscatter near the lid, leading to divergence. Similar problems have been found using the gradient or Clark model near walls \cite{Winckelmans2001,Vreman2004}, and using the NS-$\alpha$ model at high values of $\alpha^{2}$ \cite{Petersen2008}.
The outline of the paper is as follows. A description of the anisotropic subgrid model used is given in section 2. In this section condensed index notation is used to discuss the model for the sake of compactness. The numerical methods and treatment of the subgrid model are outlined in section 3.  Results from using both a mesh-based definition of $\alpha^{2}$ and an alternative definition are presented in section 4. Concluding remarks are then given in section 5.

\section{Model Formulation}
\label{formulation}
The Eulerian-averaged equations from Holm \cite{Holm1999}(section 12) are used as a starting point. In the development of these equations the instantaneous velocity is decomposed into a mean and a random fluctuation, and the averaging is applied at a fixed point \cite{Holm1999}. This is in contrast to the Lagrangian average, which is taken following a particle trajectory. We feel that the Eulerian average is more consistent with the manner in which the experimental and numerical data we are comparing with was collected. Differences between the Eulerian and Lagrangian averaged equations are discussed in Holm \cite{Holm1999}, and alternative Lagrangian-averaged equations are given in Marsden \cite{Marsden2003}. The Eulerian-averaged equations are,
 \begin{equation}
\partial_{i} \tl{u}_{i}=0, 
\end{equation}
\begin{equation}
\partial_{t} u_{i}+\tl{u}_{j}\partial_{j} u_{i}+u_{k}\partial_{i}\tl{u}_{k}=-\partial_{i} P +\nu\partial_{kk}u_{i}-\frac{1}{2}\partial_{i}\langle \xi_{k} \xi_{l}\rangle \partial_{k} \tl{u}_{m}\partial_{l}\tl{u}_{m},
\label{momHolm}
\end{equation}
where $P$ is a pressure-like variable,
\begin{equation}
P=p-\frac{1}{2}\tl{u}_{i}\tl{u}_{i}-\frac{1}{2}\langle \xi_{k}\xi_{l} \rangle \partial_{k}\tl{u}_{m}\partial_{l}\tl{u}_{m},
\label{pressureHolm}
\end{equation}
and with the following relationship between the smoothed and unsmoothed velocities,
\begin{equation}
u_{i}=\underbrace{\left(1-\partial_{k}\left(\langle \xi_{k} \xi_{l} \rangle \partial_{l}\right)\right)}_{H}\tl{u}_{i}.
\label{smvel}
\end{equation}
In \eqref{smvel} $\tl{u}$ is a smoothed velocity, $\langle \xi_{k} \xi_{l} \rangle $ is the smoothing scale, and the angle brackets, $\langle \cdot \rangle$ denote an Eulerian average. For the isotropic model $\langle \xi_{k}\xi_{l}\rangle=\alpha^{2}\delta_{kl}$.  The last term on the RHS of \eqref{momHolm} arises in the derivation when the functional derivative is taken with respect to $\langle \xi_{k}\xi_{l} \rangle$ as is necessary to conserve momentum when $\langle \xi_{k}\xi_{l} \rangle$ is not constant. The momentum equation can also be written in momentum-conservation form as \cite{Chen1999b}
\begin{equation}
\partial_{t} u_{i}+\tl{u}_{j}\partial_{j} u_{i}=-\partial_{i} p + \partial_{j}\left(\langle \xi_{k}\xi_{j} \rangle \partial_{i}\tl{u}_{m}\partial_{k}\tl{u}_{m}\right)+\nu\partial_{kk}u_{i}.
\label{NSalpha_anis}
\end{equation}
While versions of the governing equations with the smoothed velocity as the dependent variable have appeared in the literature \cite{Holm1999,Chen1999c}, the approach presented here is more familiar to the LES-community, and the purpose is to show that the subgrid stress $m_{ij}$ does not arise as an `ad-hoc'  modification to the isotropic model. To develop an equation with the smoothed velocity as the dependent variable we therefore follow the approach taken in Holm and Nadiga \cite{Holm2003} and use the commutator between the substantial derivative and the smoothing operator. For example, we would like to have a substantial derivative written entirely in terms of the smoothed velocity. This is done by rewriting the advective terms in  \eqref{NSalpha_anis} as,
\begin{equation}
\partial_{t} u_{i}+\tl{u}_{j}\partial_{j} u_{i}=[D/Dt,H]\tl{u}_{i}+H\left(\partial_{t}\tl{u}_{i}+\tl{u}_{j}\partial_{j} \tl{u}_{i}\right).
\label{advective}
\end{equation}
Here $[D/Dt,H]$ is the commutator between the material derivative and the Helmholtz operator, $H$ from \eqref{smvel}, $[D/Dt,H]\tl{u}=D/Dt(H(\tl{u}))-H(D/Dt(\tl{u}))$, where $H(\tl{u})=u$. Note that the substantial derivative is defined with the smoothed velocity,  $D/Dt=\partial_{t}+\tl{u}_{j}\partial_{j}$ in keeping with the fact that the advecting velocity is smoothed. The momentum equation \eqref{NSalpha_anis} can then be written as,
\begin{equation}
\partial_{t}\tl{u}_{i}+\tl{u}_{j}\partial_{j}\tl{u}_{i}=H^{-1}\left(\partial_{i}p+ \partial_{j}\left(\langle \xi_{k}\xi_{j} \rangle \partial_{i}\tl{u}_{m}\partial_{k}\tl{u}_{m}+\nu\partial_{kk}u_{i}\right)-[D/Dt,H]\tl{u}_{i}\right).
\end{equation}
It was found the commutator can be expressed 
\begin{equation}\begin{split}
\left[\partial_{t} +\tl{u}_{j}\partial_{j},\left(1-\partial_{k} \left( \langle \xi_{k}\xi_{l} \rangle \partial_{l} \right)\right)\right]\tl{u}_{i}=
\partial_{j}\left(\langle \xi_{k}\xi_{l} \rangle \partial_{k} \tl{u}_{i} \partial_{l} \tl{u}_{j}
+\langle \xi_{j}\xi_{l} \rangle \partial_{k} \tl{u}_{i}\partial_{l} \tl{u}_{k}\right)\\
-\partial_{j}\left(\left(\partial_{t} \langle \xi_{j}\xi_{l} \rangle+\tl{u}_{k}\partial_{k} \langle \xi_{j}\xi_{l} \rangle \right)\partial_{l} \tl{u}_{i} \right).
\label{commutator}
\end{split}\end{equation}
For constant, isotropic fluctuations, the first two terms on the RHS of \eqref{commutator} have the same form as the Leray model \cite{Geurts2006}, although we used $\partial_{i}\tl{u}_{i}=0$ in the development of the commutator so this cannot be considered as the true Leray subgrid stress. For the NS-$\alpha$ model the last two terms on the RHS of \eqref{commutator} can \textit{in theory} be neglected, because for the Eulerian-averaged equations each component of the particle displacement covariance is transported by the mean flow like a scalar,
\begin{equation}
\frac{D \langle \xi_{j}\xi_{l} \rangle}{Dt}=0.
\label{advect}
\end{equation}
This arises in the derivation of the Eulerian-averaged equations when it is assumed that all of the fluctuations are contained in the Eulerian field \cite{Holm1999}.  
\\
\\
In the context of LES modelling, this term represents the contribution to the subgrid stress due to the explicit change in filter width.  It is customary in the LES community to neglect these types of terms. For a recent discussion of this problem see van der Bos\cite{vanderBos2005} (and references therein). As is the case with all subgrid terms, these terms will lead to energy transfer between resolved and subgrid modes, and neglecting these terms therefore changes the subgrid transfer dynamics. For the equation above, this is clearly seen by considering the isotropic case, where the last two terms on \eqref{commutator} can be written as,
\begin{equation*}
\frac{\partial}{\partial x_{k}}\left(\left(\frac{D\alpha^{2}}{Dt}\right) \frac{\partial \tl{u}_{i}}{\partial x_{k} }\right).
\end{equation*}
The substantial derivative of $\alpha^{2}$ can be seen to play the role of a variable eddy viscosity. It will dissipate energy when $\alpha^{2}$ increases along a flow path and backscatter energy when it decreases. This is exactly the method suggested in the literature \cite{vanderBos2005} to model the commutation error in LES. The idea being that when a flow scale is advected into a region where the grid is coarser, it will go from being resolved to modelled, leading to dissipation, and vice versa when the grid is refined.  In practice, whether or not they can be neglected depends on the magnitude of these terms relative to the other subgrid terms. In the one-dimensional case they can be neglected if
\begin{equation*}
\frac{1}{\alpha^{2}} \frac{\partial \alpha^{2}}{\partial x} \ll \frac{1}{\tl{u}}\frac{\partial \tl{u}}{\partial x},
\end{equation*}
which means that the filter field must be smoother than the flow field in the direction of advection. We found that the commutation term tends to be large when the advection is large, and does not have a significant effect on the flow, as compared to the tilting term (which is $u_{k}\partial_{i}\tl{u}_{k}$ in (\ref{NSalpha_iso}) and will become part of the $\partial_{j}(C_{ij})$ term in the following). 
\\
\\
The second simplification will be to retain only the diagonal components of $\langle \xi_{k}\xi_{l} \rangle$. This is equivalent to using the Helmholtz operator, 
\begin{equation}
H=\left(1-\partial_{x}(\langle \xi_{x} \xi_{x} \rangle \partial_{x})-\partial_{y}(\langle \xi_{y}\xi_{y} \rangle \partial_{y})-\partial_{z}(\langle \xi_{z}\xi_{z} \rangle \partial_{z})\right).
\end{equation}
This gives us a formulation similar to that derived using second order reconstruction methods \cite{Winckelmans2001}, where the lack of off-diagonal terms arises when the three-dimensional filter is applied as the composition of three one-dimensional filters, $L=l_{1} \circ l_{2} \circ l_{3}$, where  $l_{j}$ ($j=1,2,3$) represents a one dimensional filter in the $x_{j}$-direction.  The benefit of such a simplification is that substantially reduces the cost and yields a subgrid model that is no more expensive than the isotropic version (which is itself about as costly as the dynamic model \cite{Geurts2006}). 
\\
\\
With the above assumptions and simplifications, the momentum equation can then be written
\begin{equation}
\partial_{t}\tl{u}_{i}+\partial_{j}\tl{u}_{i}\tl{u}_{j}=-\partial_{i}\tl{p}+\nu\partial_{kk}\tl{u}_{i}-H^{-1}(\partial_{j}m_{ij}).
\end{equation}
The subgrid stress is
\begin{equation}
m_{ij}=
\alpha^{2}_{k}\delta_{kl} \frac{\partial \tl{u}_{i}}{\partial x_{k}}\frac{\partial \tl{u}_{j}}{\partial x_{l}}
+\alpha^{2}_{l}\delta_{jl} \frac{\partial \tl{u}_{k}}{\partial x_{l}}\frac{\partial \tl{u}_{i}}{\partial x_{k}}
-\alpha^{2}_{k}\delta_{kj} \frac{\partial \tl{u}_{m}}{\partial x_{i}}\frac{\partial \tl{u}_{m}}{\partial x_{k}},
\label{sgsstress}
\end{equation}
where we have used $\alpha_{k}^{2}$ for $\langle \xi_{k}\xi_{k}\rangle$. Following Geurts and Holm \cite{Geurts2006} the subgrid model can also be written as
\begin{equation}
m_{ij}=A_{ij}+B_{ij}-C_{ij}.
\end{equation}
Here, $A_{ij}$ is the anisotropic gradient model (or Clark model), $A_{ij}+B_{ij}$ is similar to a Leray model and the NS-$\alpha$ model is comprised of all three terms. 

\subsection{Physical Interpretation of the Subgrid Term}
\label{sgsterm}
When written as a subgrid stress the effect of the $m_{ij}$ term is not easily interpreted physically. This term provides force to the momentum equation, and we will now rewrite it such that the form of this forcing function is clarified. Complementary discussions along these lines have been given in the literature \cite{Holm1999,Domaradzki2001}, here we mention this explicitly because of its relevance to the following section. To keep things simple in this section we will assume isotropic fluctuations, and that $\alpha^{2}$ is constant. We will also make use of the difference between the smoothed and unsmoothed velocity, 
\begin{equation*}
u^{ST}_{i}=\alpha^{2}\frac{\partial^{2}\tl{u}_{i}}{\partial x_{k}^{2}}.
\end{equation*}
In the LES literature this velocity would be called the subgrid fluctuation. In the NS-$\alpha$ literature it is referred to \cite{Holm1999} as the `Stokes velocity'. When $\alpha^{2}$ is constant the commutator can be written using the Stokes velocity as, 
\begin{equation}
\left[ D/Dt,H \right] \tl{u}_{i}=2\partial_{j}A_{ij}-u_{j}^{ST}\frac{\partial \tl{u}_{i}}{\partial x_{j}}.
\label{CLmom}
\end{equation}
and the  momentum equation can be written,
\begin{equation}
\partial_{t} \tl{u}_{i}+\tl{u}_{j}\partial_{j} \tl{u}_{i}=-\partial_{i} \tl{p}^{*}+\nu\partial_{kk}\tl{u}_{i}-H^{-1}\left(2 \partial_{j} A_{ij}-u^{ST}\times \omega \right)
\end{equation}
where $p^{*}$ is a modified pressure.
The term $u^{ST} \times \tl{\omega}$ in \eqref{CLmom} is called the vortex force. We can now see that the subgrid model is composed of two forcing terms. The first term, $A_{ij}$, is well known in the literature where it goes by many names, such as the Clark model \cite{Graham2008}, gradient model and Tensor-Diffusivity model \cite{Vreman1996,Winckelmans2001}.   It  is a generic subgrid closure which can be derived by expanding the subgrid stress $\tau_{ij}$ in a Taylor series expansion and retaining terms up to $O(\Delta^{2})$, where $\Delta$ is the filter width. When the Helmholtz operator is approximated by a box filter, $\alpha^{2} \sim \Delta^{2}/24$, where $\Delta$ is filter width, this term is identical to the model in the literature \cite{Winckelmans2001}
\begin{equation*}
2A_{ij}=2\alpha^{2}\frac{\partial \tl{u}_{i}}{\partial x_{k}}\frac{\partial \tl{u}_{j}}{\partial x_{k}}\\
       =\frac{\Delta^{2}}{12}\frac{\partial \tl{u}_{i}}{\partial x_{k}}\frac{\partial \tl{u}_{j}}{\partial x_{k}}.
\end{equation*}
This suggests an alternative (more approximate) way of deriving an equation for the smoothed velocity, which would be to start with the momentum equation, rewrite it with the Stokes vortex force on the RHS, apply a filter to the equations, and close the resulting $\tau_{ij}$ term with an explicitly filtered gradient model. 
\\
\\
The vortex force is what makes the NS-$\alpha$ model different from other approaches. To highlight how a vortex forcing term is fundamentally different than, for example, a Smagorinsky model, consider a simple two-dimensional mixing layer with $u=\tanh(y)$. The Smagorinsky model will add a diffusion term to the momentum equation, with diffusivity that is a function of the filtered strain-rate $\tl{S}_{ij}$,
\begin{equation*}
\nu_{T}=\left(C_{s}\Delta \right)^{2} \left(2 \tl{S}_{ij} \tl{S}_{ij}\right)^{1/2}\sim \left(C_{s}\Delta \right)^{2} {\rm sech} ^{2}y.
\end{equation*}
The diffusivity will be highest at the middle of the mixing layer, and it is not surprising that such a model cannot be used for studies of mixing layer transition, where it damps out the small amplitude perturbations preventing transition. 
\\
\\
On the other hand, the vortex force would make its most significant contribution to the vertical ($y$) momentum equation, with a vertical forcing term
\begin{equation*}
u^{ST}\omega_{z} \sim \alpha^{2}_{y}\frac{\partial^{2}\tl{u}}{\partial y^{2}}\omega_{z}.
\end{equation*}
At the very early stages of transition this term would provide equal and opposite vertical forcing to the mixing layer, and therefore leaves the mixing layer unchanged. However, as soon as undulations in the layer appear the flow is no longer symmetric and such a terms would serve to `push'  the mixing layer back and forth. Unlike the Smagorinsky model, the NS-$\alpha$ model was found to correctly capture the linear growth phase of a transitional mixing layer \cite{Geurts2006}

\section{Numerical Methods}
\label{nummeth}
The governing equations for $\tl{u}$ are solved using the STREAM code \cite{Lien1994a}. This is a collocated finite-volume code which uses the SIMPLEC method to ensure mass conservation. The advection and diffusion terms are treated implicitly while the $m_{ij}$ term is treated using deferred correction. A second-order time stepping scheme is used with $CFL \approx 1$. The advection scheme used was the QUICK scheme. Note that this is very similar to methods used in earlier studies of this flow \cite{Freitas1988} and more recently in the evaluation of the dynamic mixed model on the same test case \cite{Zang1993}.   
\\
\\
In the finite volume formulation the source term appears on the right-hand side of the
momentum equation as
\begin{equation}
\int_{V}H^{-1}\left(\frac{\partial m_{ij}}{\partial x_{j}}\right)\,dV.
\end{equation}
If we first calculate the $m_{ij}$ term at cell centers and then the divergence, this means the velocity gradients in the $m_{ij}$ terms would be computed at cell centers. It was found that doing this using second order central differences made the model even more susceptible to the numerical oscillations we will see in the next section. As an alternative, the method used here is to write the source term in a manner consistent with the other terms in the momentum equation
\begin{equation}
H^{-1}\int_{V}\frac{\partial}{\partial x_{j}}\left(m_{ij}\right) \, dV \simeq H^{-1}\int_{CS}m_{ij}A_{j}. 
\end{equation}
The velocity gradients are now computed at control volume faces, which means the source term is computed from velocity differences between adjacent nodes. Thus the procedure is to first compute the subgrid force at all interior nodes, 
\begin{equation}
F_{i}=\int_{CS}m_{ij}A_{j},
\end{equation}
and then find the filtered force $\tl{F}_{i}$. For the present results this was done by solving the Helmholtz equation
\begin{equation}
F_{i}=\tl{F}_{i}-\frac{\partial}{\partial x_{k}}\left (\alpha_{k}^{2}\frac{\partial \tl{F}_{i}}{\partial x_{k}}\right),
\end{equation}
using a conjugate gradient solver. The solution of the Helmholtz equation requires boundary conditions for $\tl{F}_{i}$, but in the discrete equation the boundary value of $\tl{F}_{i}$ will be multiplied with a boundary value of $\alpha^{2}_{k}$, which we know is zero from the boundary condition $\langle \xi_{k}\xi_{l} \rangle \hat{n}_{k}=0$. Simulations were also carried out using a simple box filter, as has been done in previous studies \cite{Geurts2006}. This was found be be much more efficient than Helmholtz inversion, with comparable results.

\section{Results}
\label{results}

\subsection{Description of the Test Case}
\label{testcase}
\begin{figure}
\centering
\epsfig{file=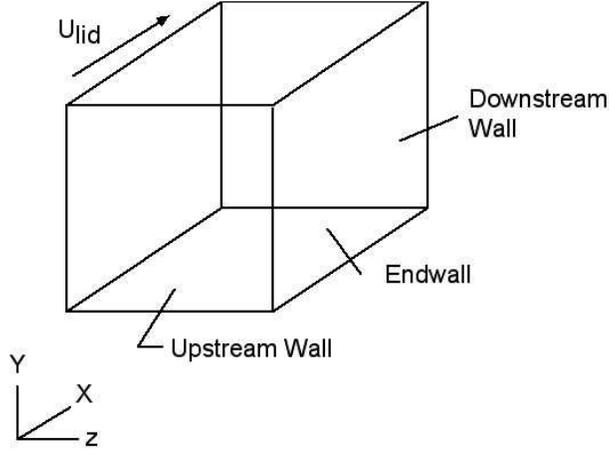,height=6cm}
\caption{Sketch of the lid-driven cavity flow.}
\label{cavitysketch}
\end{figure}

The application of the NS-$\alpha$ model to a practical problem is studied here using a lid-driven cavity flow at a Reynolds number of 10,000, where the Reynolds number is based on the lid velocity and cavity length. The chosen cavity has a spanwise aspect ratio (SAR) of 1, as shown in the schematic in Figure \ref{cavitysketch}. The three-dimensional cavity flow contains a variety of flow structures and is a challenging test case for a subgrid model due to the lack of homogeneous directions, the presence of both laminar and turbulent flow regions and the anisotropic nature of the flow. This cavity has been studied both experimentally \cite{Prasad1989} and numerically using LES \cite{Zang1993,Bouffanais2007} and DNS \cite{Leriche2000}. At this Reynolds number the distinguishing feature of the flow is the formation of two jets which  separate off the downstream wall and impinge on the cavity bottom. While the experimental measurements reported a small inertial subrange near the cavity bottom, the DNS study reports that the flow does not actually become fully turbulent before it encounters the upstream wall. There are however,  significant regions of both positive and negative turbulent energy production in the jet impingement regions, as is discussed in detail in the DNS paper \cite{Leriche2000}. We would like to see to what extent the model investigated here can capture this.
\\
\\
The mesh used is stretched in the $x$ and $y$ directions to capture the shear layers near the walls, but uniform in the spanwise since the relevant flow physics are not clustered near the endwalls, but distributed along the span. The parameters pertaining to the mesh sizes and stretching ratios are given in Table 1. The meshes are similar to those used in the study of Zang et al. \cite{Zang1993} using the dynamic mixed model where similar numerical methods were employed. 
\\
\\
To assess the time step  we first compare our parameters to those used in the experiments. For the lid-driven cavity all quantities are non-dimensionalized by the cavity length $(L)$ and lid velocity $(U)$. The characteristic time scale is then $L/U$ which can be written in terms of the Reynolds number as,
$\frac{L}{U}=Re\frac{L^{2}}{\nu}$.
Estimating the kinematic viscosity of water at room temperature as $1 \times 10^{-6} m^{2}/s$ and knowing the length of the cavity to be $150mm$ gives $L/U=2.25s$. A time step of $0.01$ then corresponds to physical time of $0.025s$ or physical frequency of $40Hz$. The power spectra shown from the experiments for all cases have very little frequency content above $1Hz$. Therefore, it was expected that the timestep chosen would be adequate. This  was verified by the fact that simulation results showed very little difference when run at a timestep half as big. Rolling averages of the mean and rms velocities, as well as the total kinetic energy and dissipation were monitored. After a statistically steady state was achieved, statistics were collected over $40 L/U$.

\subsection{Results with $\alpha_{k}^{2}$ based on the mesh}
\label{resultsmesh}

\begin{figure}
\centering
\psfrag{U}{$\langle \tl{u} \rangle$}
\psfrag{V}{$\langle \tl{v} \rangle$}
\epsfig{file=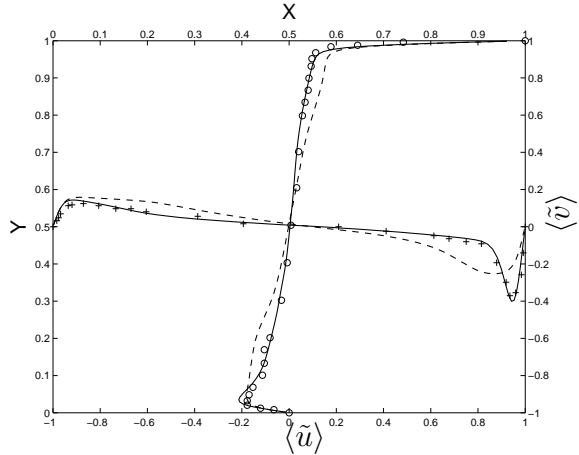,height=6cm}
\caption{Mean flow profiles on the midplane for the $(64)^{3}$ mesh showing the wall jet is pushed out too far from the downstream wall when mesh-based $\alpha^{2}_{k}$ is used. Solid line, no model; dashed line, NS-$\alpha$ model with $\alpha^{2}_{k}$ based on the mesh. Symbols
are experimental data \cite{Prasad1989}.}
\label{highRe}
\end{figure}

\begin{figure}
\centering
\psfrag{urms}{$10 u_{rms}$}
\psfrag{vrms}{$10 v_{rms}$}
\mbox{
\subfigure[$\alpha^{2}_{z}$ based on the mesh ]{\epsfig{file=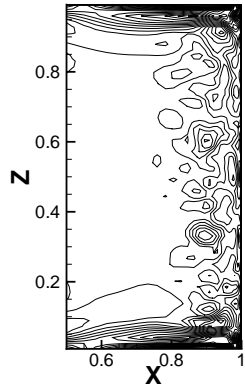,height=6cm}}}
\mbox{
\subfigure[$\alpha^{2}_{z}=0$ ]{\epsfig{file=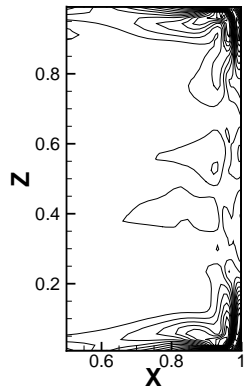,height=6cm}}}
\caption{Vertical vorticity $\tl{\omega}_{y}$ near the downstream wall demonstrating the small-scale vorticity found due to the $\alpha^{2}_{k}$ discontinuity. The plane is at a height of $y=0.6$.}
\label{verticalvortices}
\end{figure}

\begin{figure}
\epsfig{file=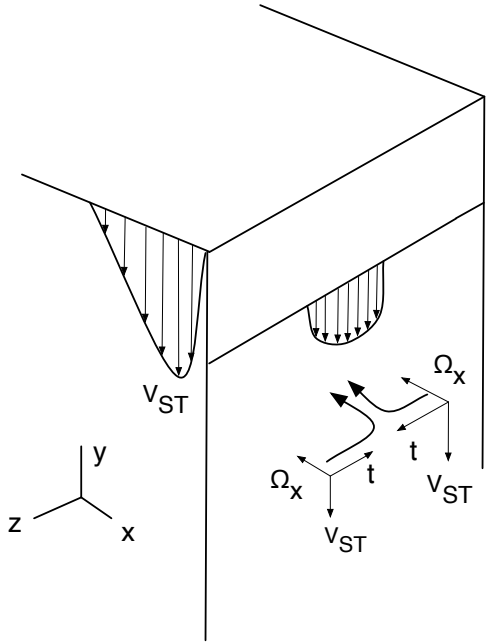,height=9cm}
\caption{A spanwise perturbation generates vertical vorticity. Opposing torques $\boldsymbol{t}$ are created from the $\boldsymbol{u}^{ST}\times \boldsymbol{\omega}$ term in the momentum equation. These create a convergence zone and vertical vorticity. In the vorticity equation, horizontal vorticity $(\tl{\omega}_{x})$ is tilted by $v^{ST}$ into the vertical direction. After the sketch in Leibovich \cite{Leibovich1983}, rotated $90^{o}$.}
\label{CLmech}
\end{figure}

Since $\alpha_{k}^{2}$ is a smoothing scale we start with a simple definition based on the grid size 
\begin{equation}
\alpha^{2}_{k}=C\left(h_{k}^{2}\right)
\label{alphamesh}
\end{equation}
where $h_{k}$ is the grid spacing in the \textit{k}-direction and $C$ is a constant denoting what fraction of the grid spacing to use.  Because $\alpha_{k}^{2}$ can be related to the filter width, $\Delta_{k}$, of a box filter via $\alpha_{k}^{2}=\Delta^{2}_{k}/24$ \cite{Geurts2003}, we choose $C=1/6$, which corresponds to a filter width which is twice the grid size. It has been suggested that for adequate subgrid resolution in the isotropic model $\alpha^{2}=h^{2}$ should be used for the NS-$\alpha$ model \cite{Geurts2006}. In the present study the problems encountered were in laminar flow regions, where the model should be inactive, and the question of subgrid resolution was not addressed in detail. However, in some cases simulations were run on different meshes to verify the sensitivity of the results to the observed trends. In all cases the value of $C$ was adjusted so that the physical value of $\alpha_{k}^{2}$ was approximately the same as on the coarse mesh. Simulations were also done with the isotropic model with $\alpha^{2}$ proportional to the volume, but as mentioned in the introduction, this model was found to be generate excessive backscatter near the cavity lid, leading to divergence.
\\
\\
It was found that there was a very persistent problem when $\alpha^{2}_{k}$ was based on the mesh. This was that the downstream wall jet was pushed too far out from the wall, as shown in Figure \ref{highRe}. This was observed on both coarse and refined meshes, over a range of $\alpha_{k}^{2}$ values, and also when a box filter was used instead of a Helmholtz operator. It was  also seen when the isotropic version of the model (with $\alpha^{2}$ based on the grid volume) was used. 
\\
\\
Because the wall jet is pushed too far out from the downstream wall (in the $x$-direction) we now look at the contribution of the subgrid force to the $\tl{u}$-momentum equation. Using the vortex force,
\begin{equation*}
F_{x}=\partial_{j}A_{1j}-\left(v^{ST}\tl{\omega}_{z}-w^{ST}\tl{\omega}_{y}\right).
\end{equation*}
In the wall jet region we found the vortex forcing term is much larger than the $A_{ij}$ term. In this region, $\tl{\omega}_{z} \gg \tl{\omega}_{y}$ and
$\partial_{x} \gg \partial_{y},\partial_{z}$ and it was expected that the
problem was coming from following component of the vortex force,
\begin{equation*}
\alpha_{x}^{2}\frac{\partial^{2} \tl{v}}{\partial x^{2}}\tl{\omega}_{z}.
\end{equation*}
Given that the vorticity field is unsteady this will be an unsteady
forcing term which could cause the wall jet to oscillate back and forth,
leading to high fluctuation levels. Depending on the balance between the
positive and negative forcing, it is also possible that this could
lead to the jet being pushed too far out from the wall in the mean. For
the anisotropic model this hypothesis could be easily tested by turning
off $\alpha^{2}_{x}$. To our surprise this did not help the situation. Instead, it
was turning off $\alpha^{2}_{z}$ which solved this problem.
\\
\\
A closer examination of the flow fields corresponding to the $\alpha^{2}_{z}
\ne 0$ and $\alpha^{2}_{z} =0$ cases showed that the main difference between
the two is the appearance of streamwise (here vertical) vortices ($\tl{\omega}_{y}$) in the downstream walljet region when $\alpha^{2}_{z} \ne 0$, as shown in Figure \ref{verticalvortices}.  These vortices do not appear when a model is not used and appear to be 
a numerical artifact of the NS-$\alpha$ model. These
presence of these vortices can be understood if there is significant
modulation of the velocity in the spanwise direction, as for example
could be caused by spanwise numerical oscillations. Recall that the
vortex force in the momentum equations appears as advection and
stretching/tilting terms in the vorticity equations. In particular the
stretching/tilting term in the vertical vorticity equation is,
\begin{equation*}
\tl{\omega}_{x}\frac{\partial v^{ST}}{\partial
x}+\tl{\omega}_{y}\frac{\partial v^{ST}}{\partial
y}+\tl{\omega}_{z}\frac{\partial v^{ST}}{\partial z}.
\end{equation*}
Oscillations in the spanwise direction would lead to the generation of streamwise vorticity, tilted into the vertical by the first term.  It was found that the flat velocity profile
seen in Figure \ref{highRe} developed slowly over time, indicating that the
long-time average effect of the vertical vortices is a mixing out of the
velocity profile. Similar problems were observed in cases where the
numerical oscillations were not as visually obvious (at a lower Reynolds number of $Re=3,200$), again showing that since the effect builds up
over time, small oscillations can have a significant impact. 
\\
\\
There is an interesting analogy between the behavior seen here, and the true physical behavior of the NS-$\alpha$ equations. It has been pointed out \cite{Holm1999} that these equations are related to the  Craik-Leibovich equations \cite{Craik1976}, which are used to account for the long-time averaged effect of surface waves on the background current.  The effect of surface waves is to create a relative velocity between a fluid particle (Lagrangian) and the background current (Eulerian). This relative velocity, which is called the Stokes drift velocity, then tilts vertical vorticity into the streamwise direction to create streamwise vortices (Langmuir cells) which transport momentum
perpendicular to the free surface and flatten the velocity profile below
the surface, leading to a mixed layer \cite{McWilliams1997}, a schematic is shown in Figure \ref{CLmech}.
Thus, this result is not erroneous in the sense
that it is a real solution to the given equations in the presence of
small scale spanwise oscillations. However, the problem is that the
oscillations are not coming from something physical, such as surface
waves, but from an unwanted numerical effect. 
\\
\\
To verify that this is not just a boundary problem a simple test was done with a laminar Couette flow and an $\alpha^{2}_{y}$ discontinuity was introduced in the middle. Once again, the model generated a force due to the $C_{22}$ term, which was balanced by a pressure gradient. This was tested with varying subgrid resolutions (i.e. keeping the physical size of $\alpha^{2}_{y}$ constant but refining the mesh) and was found to be relatively insensitive to the resolution. 
\\
\\
Although this is not a boundary problem per se, it can be readily verified that if both $\alpha^{2}_{y}\rightarrow 0$ and $\partial_{y} \alpha^{2}_{y}\rightarrow 0$ at the lid, we would not have this problem. From the relationship between the smoothed and unsmoothed velocity fields,
\begin{align}
u &=\tl{u}-\frac{\partial}{\partial y}\left(\alpha^{2}_{y}\frac{\partial \tl{u}}{\partial y}\right),\\
u &=\tl{u}-\frac{\partial \alpha^{2}_{y}}{\partial y}\frac{\partial \tl{u}}{\partial y} -\alpha^{2}_{y}\frac{\partial^{2}\tl{u}}{\partial y^{2}},
\label{smunsm}
\end{align}
\noindent
we can see this corresponds to both fields satisfying the same boundary condition, which here would be the no-slip condition. Unfortunately when $\alpha_{y}^{2}$ is based on the mesh and the mesh is uniform it is impossible to satisfy both $\alpha_{y}^{2}\rightarrow 0$ and $\partial_{y}\alpha_{y}^{2}\rightarrow 0$. Note that the same problem arises for the isotropic version of the model, which was also verified numerically. 

\subsection{An alternative definition of $\alpha_{k}^{2}$}
\label{alphadefn}

As an alternative to having $\alpha_{k}^{2}$ based strictly on the mesh spacing, we have used a definition of $\alpha_{k}^{2}$ which incorporates the properties of the resolved flow. There are a number of ways one could do this, and the definition given here is one which we found to work well.  
\\
\\
To gain further insight in how to define $\alpha_{k}^{2}$ we can go back to Taylor's paper \cite{Taylor1922} on turbulent diffusion since, in theory, $\alpha_{k}^{2}$ is a measure of mean-squared particle displacement. The relationship between the NS-$\alpha$ model and Taylor's work on turbulent diffusion has been discussed earlier \cite{Holm2005}. First of all, Taylor showed that if the averaging time for the particle motion $(T)$ is long relative to the time over which the particle takes a step $(\tau)$ the scaling of the mean-squared particle displacement will be, $[ X^{2} ]  \sim [ v ] ^{2}T^{2}$, where $[ v ] ^{2}$ is a measure of the particle velocity and $[ \cdot ]$ denotes an ensemble average. In contrast if the averaging time is short relative to the step time $T\sim \tau$ the scaling will be, $[ X ^{2}]  \sim [ v ] ^{2} T \tau$. In the development of the NS-$\alpha$ model it is assumed that there is a separation of scales \cite{Holm1999}. Thus we will apply the former scaling here and use a time scale based on the velocity gradient tensor $T^{2}\sim \tl{g}_{ij}\tl{g}_{ij}$ where $\tl{g}_{ij}=\partial_{j}\tl{u}_{i}$. To form  a velocity scale we again follow Taylor \cite{Taylor1922} who pointed out that in considering the dispersion of a particle due to turbulent motion it is not the kinetic energy of the particle $v^{2}$ that is relevant, but the number of times it changes direction. In one dimension this can be captured by $\left(\partial_{x} v\right)^{2}$ or $\left(\partial_{t} v \right)^{2}$. In the more general case a second-order structure function could be used. In the anisotropic case this would be \cite{Lesieur1996},  
\begin{equation}
F_{2}(\mathbf{x},\Delta,t)=\frac{1}{6}\sum_{i=1}^{3}
[ || \mathbf{u}(\mathbf{x},t)-\mathbf{u}(\mathbf{x}+\Delta x_{i}\mathbf{e}_{i},t)||^{2}-
  || \mathbf{u}(\mathbf{x},t)-\mathbf{u}(\mathbf{x}-\Delta x_{i}\mathbf{e}_{i},t)||^{2}]
  \left(\frac{\Delta}{\Delta x_{i}}\right)^{2/3}.
\end{equation}
Here $\mathbf{e}_{i}$ denotes a unit vector and $\Delta$ is a length scale based on the grid volume as $\Delta = \left(h_{1}h_{2}h_{3}\right)^{1/3}$. For homogeneous, isotropic turbulence this is similar to using the turbulent kinetic energy to estimate $[v]^{2}$ since in that case there is a simple relationship between the second order structure function and the energy spectral density (see Batchelor p. 120 \cite{Batchelor1972}). 
\\
\\
Putting the velocity and time scales together we would then arrive
at the following definition for $\alpha_{k}^{2}$, 
\begin{equation}
\alpha_{k}^{2}=
\frac{F_{2}(\mathbf{x},\Delta,t)}
{\tl{g}_{ij}\tl{g}_{ij}}.
\end{equation}
In practice $F_{2}$ is computed using the six closest neighbors to a given mesh point \cite{Ducros1996}. This means such a definition of $\alpha_{k}^{2}$ would reduce to the wall normal spacing  in a wall-bounded flow, which will result in little improvement over the simple grid-based definition. This problem can be anticipated because in a wall bounded flow, for example a channel flow with $\partial \tl{u}/\partial y$ as the shear, the velocity fluctuation associated with $\tl{u}(y+\Delta y)-\tl{u}(y)$ is not fully turbulent, and should not be included in the computation of $F_{2}$. This problem has been discussed in the literature in applications of the structure function model to channel and boundary layer flows \cite{Ducros1996}. In this case the problem was resolved by not including $\tl{u}(y+\Delta y)-\tl{u}(y)$ in the calculation of $F_{2}$. In the more complex situation other strategies, such as high pass filtering, are often used \cite{Ducros1996}. 
\\
\\
The definition which was found to work well instead was,
\begin{align}
\alpha_{x}^{2}&=\max \left[(\delta_{x} \tl{u})^{2},(\delta_{y} \tl{u})^{2},(\delta_{z}
\tl{u})^{2}\right] \, T^{2}
\label{alphataylor1}
\\
\alpha_{y}^{2}&=\max \left[(\delta_{x} \tl{v})^{2},(\delta_{y} \tl{v})^{2},(\delta_{z}
\tl{v})^{2}\right] \, T^{2}
\label{alphataylor2}
\\
\alpha_{z}^{2}&=\max \left[(\delta_{x} \tl{w})^{2},(\delta_{y} \tl{w})^{2},(\delta_{z}
\tl{w})^{2}\right] \, T^{2}
\label{alphataylor3}
\end{align}
where again $T^{2}$ is $(\tl{g}_{ij}\tl{g}_{ij})^{-1}$ and the $\delta$ symbol denotes a velocity increment. In practice this can computed as the velocity difference between adjacent mesh points.
Whereas a structure function is based on the velocity difference in a
given direction and tells us about energy contained in eddies of a
given size, this definition tells us about the energy in the horizontal,
vertical and spanwise velocity fluctuations.  The question then arises as to which is more appropriate. The
definition given above was based on heuristic reasoning. If a
blob of fluid is experiencing an oscillating shear force, it would be
the $\partial \tl{u}/\partial y$ shear which would cause it to move back
and forth in the horizontal direction, while the $\partial
\tl{v}/\partial x$ shear would cause it move back and forth in the
vertical direction. Thus it was reasoned that $\alpha^{2}_{x}$ should be
related to $\delta \tl{u}_{y}$ and not $\delta \tl{v}_{x}$. 

\subsection{Results from the alternative definition}
\label{resultsalt}

\begin{figure}
\mbox{
\subfigure[\tiny Without model]{
\epsfig{file=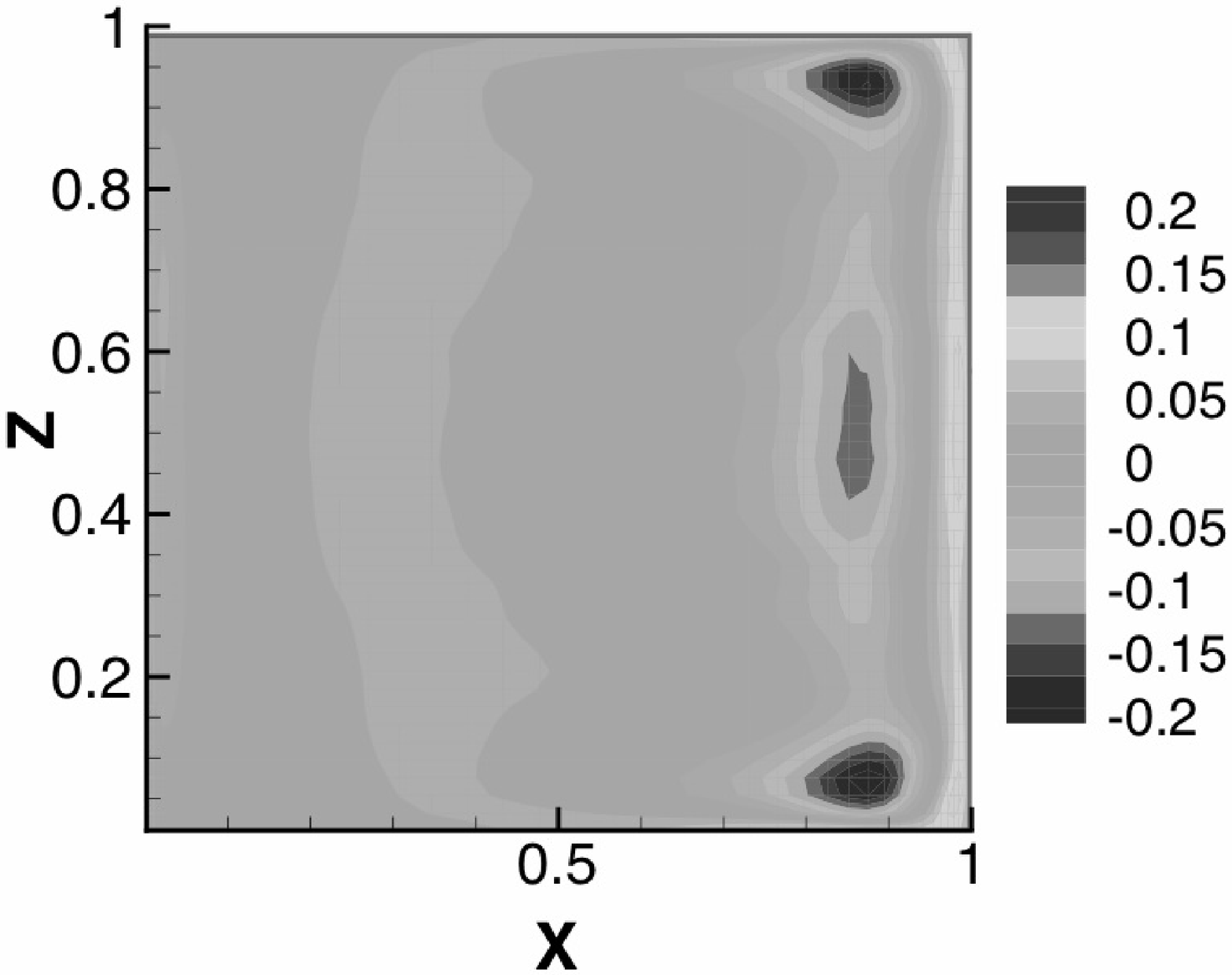,height=6.cm}}}
\mbox{
\subfigure[\tiny With model]{
\epsfig{file=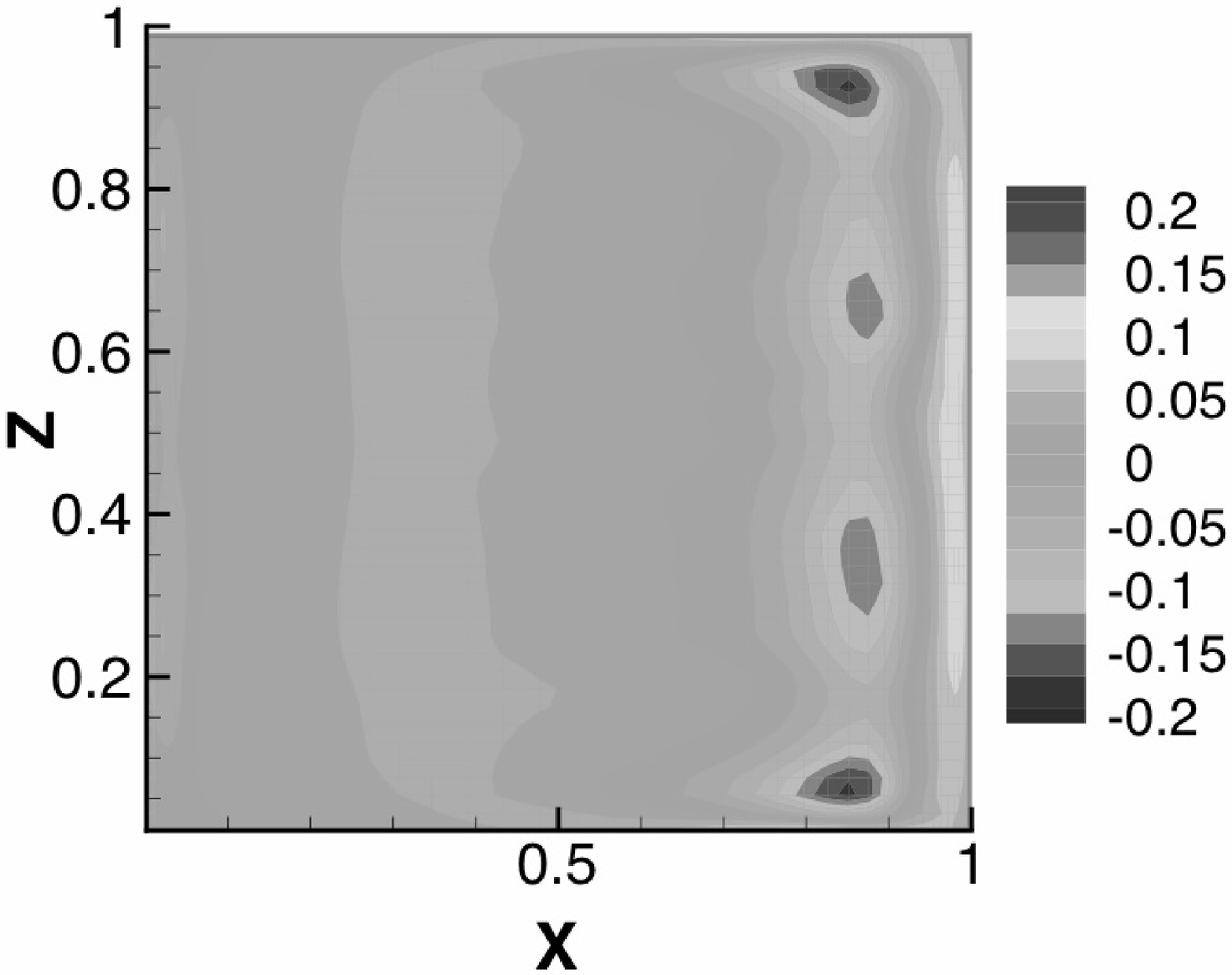,height=6.cm}}}
\caption{Contours of $\langle \tl{v} \rangle$ looking down at the cavity bottom showing that the wall jet correctly splits into two when the model is used, but does not split without the model.}
\label{walljets}
\end{figure}

\begin{figure}
\centering
\mbox{
\psfrag{U}{$\langle \tl{u} \rangle$}
\psfrag{V}{$\langle \tl{v} \rangle$}
\subfigure[Mean flow ]{\epsfig{file=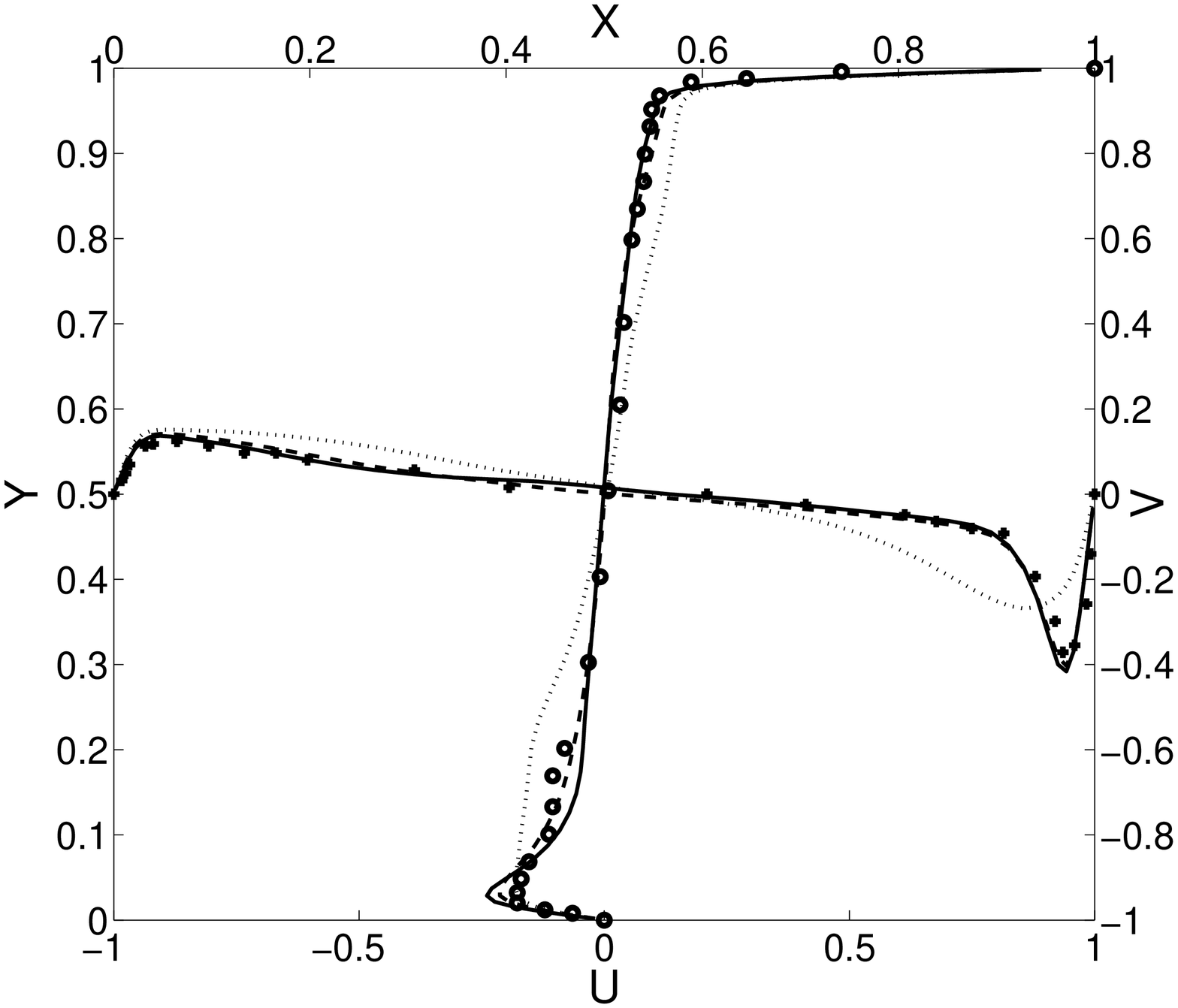,width=7cm}}}
\hfill
\psfrag{urms}{$10 u_{rms}$}
\psfrag{vrms}{$10 v_{rms}$}
\mbox{
\subfigure[rms profiles ]{\epsfig{file=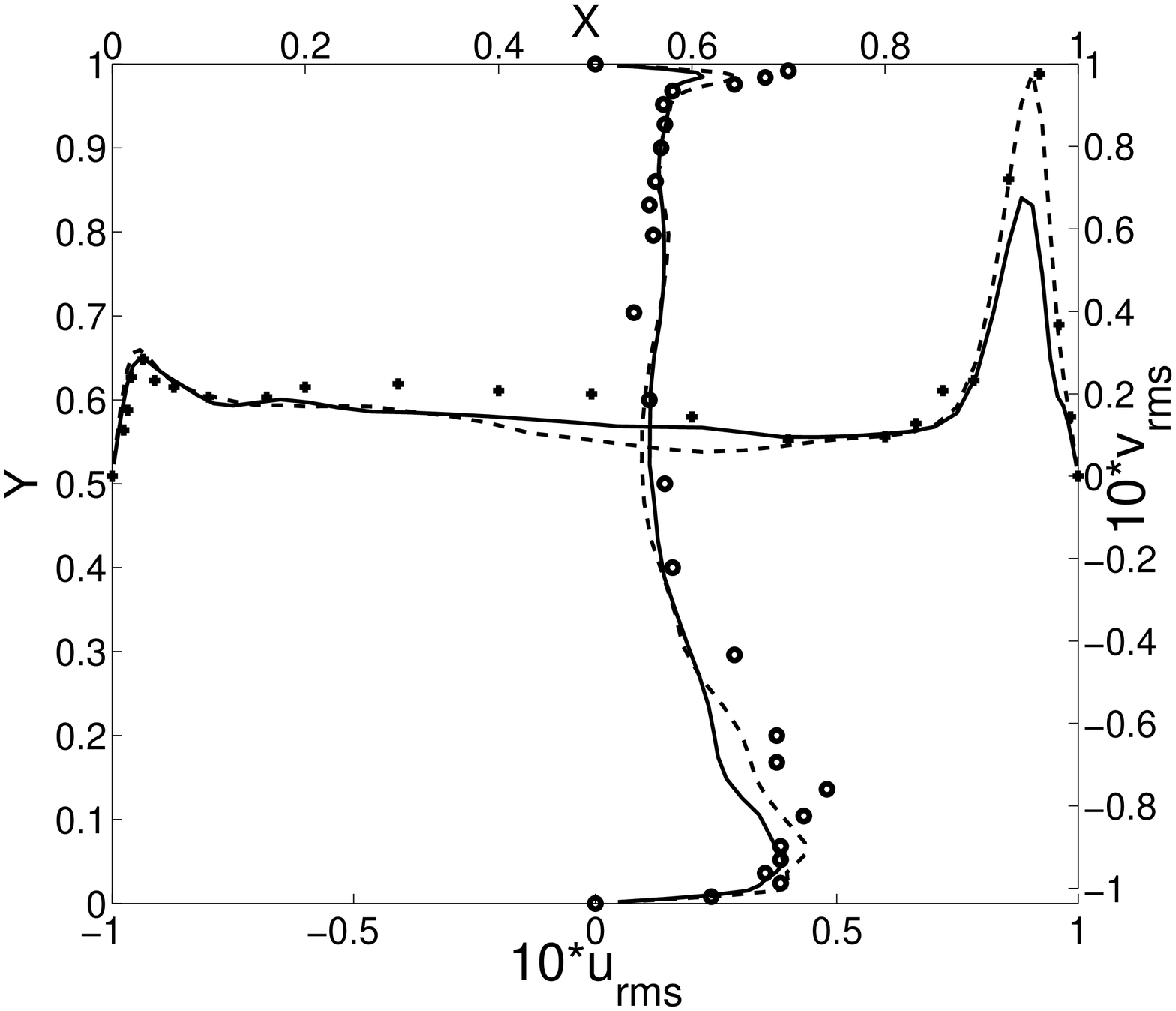,width=7cm}}}
\mbox{
\psfrag{uv}{$500\langle \tl{u}' \tl{v}' \rangle$}
\subfigure[$\langle \tl{u}' \tl{v}' \rangle$ profiles ]{\epsfig{file=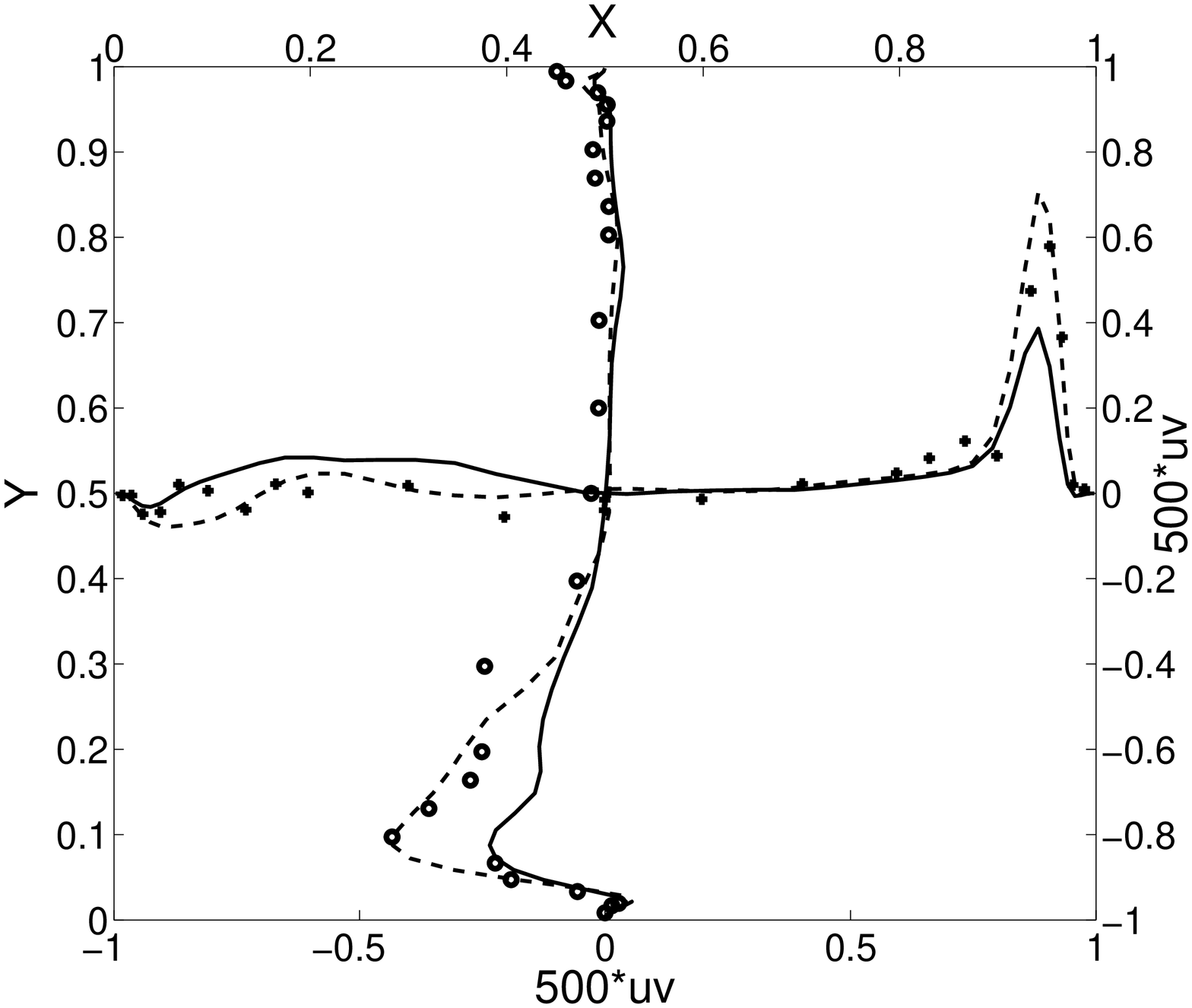,width=7cm}}}
\caption{Mean flow, rms and $\langle \tl{u}'\tl{v}' \rangle$ profiles on the midplane for the $48^{3}$ mesh. Solid line is no model, dotted line (first plot only) is NS-$\alpha$ with the mesh-based definition of $\alpha^{2}_{k}$ and dashed line is NS-$\alpha$ with alternative definition of $\alpha_{k}^{2}$. Symbols are experimental data.\cite{Prasad1989}}
\label{profiles48}
\end{figure}

\begin{figure}
\centering
\mbox{
\psfrag{U}{$\langle \tl{u} \rangle$}
\psfrag{V}{$\langle \tl{v} \rangle$}
\subfigure[Mean flow ]{\epsfig{file=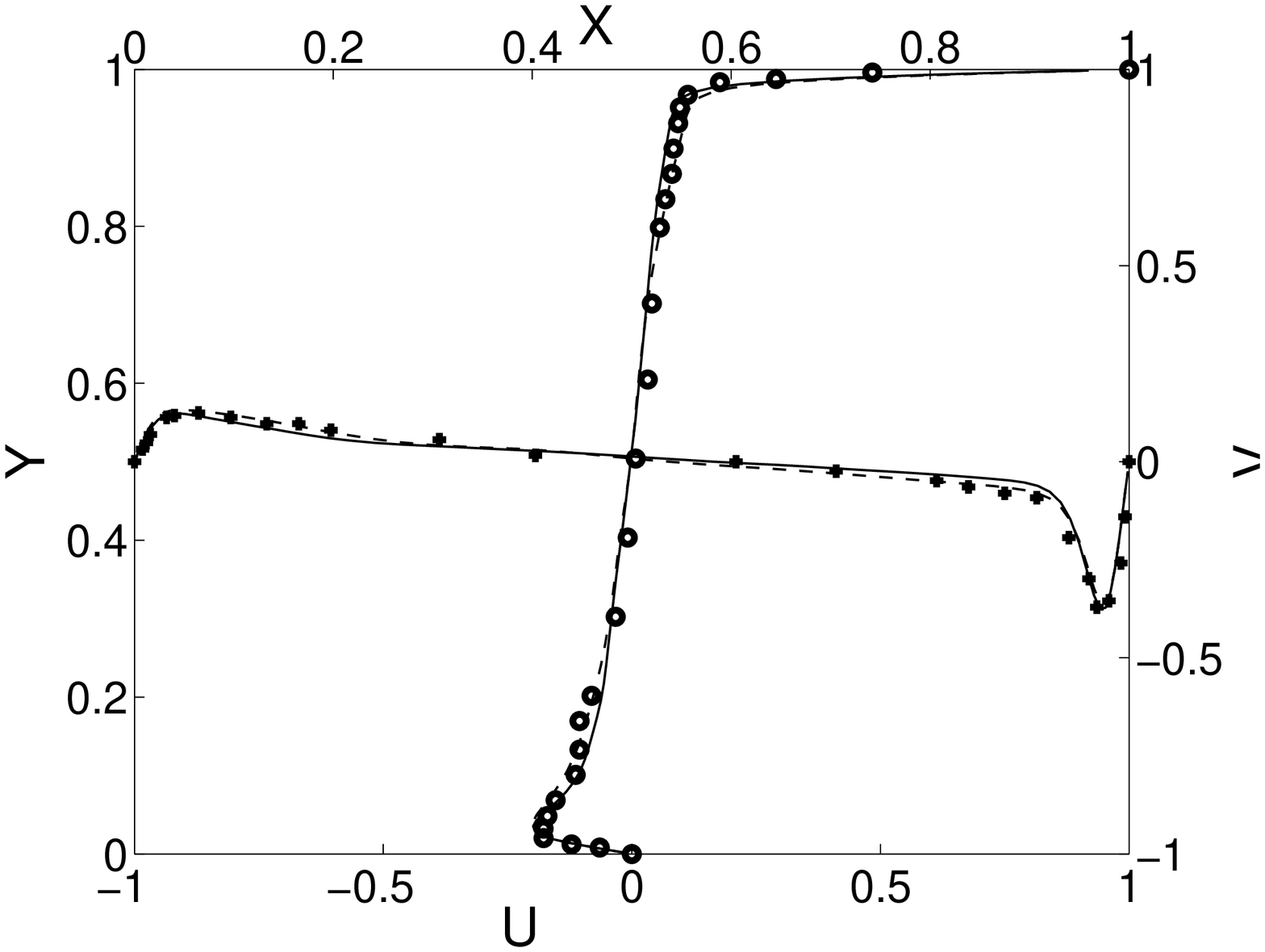,width=7cm}}}
\hfill
\psfrag{urms}{$10 u_{rms}$}
\psfrag{vrms}{$10 v_{rms}$}
\mbox{
\subfigure[rms profiles ]{\epsfig{file=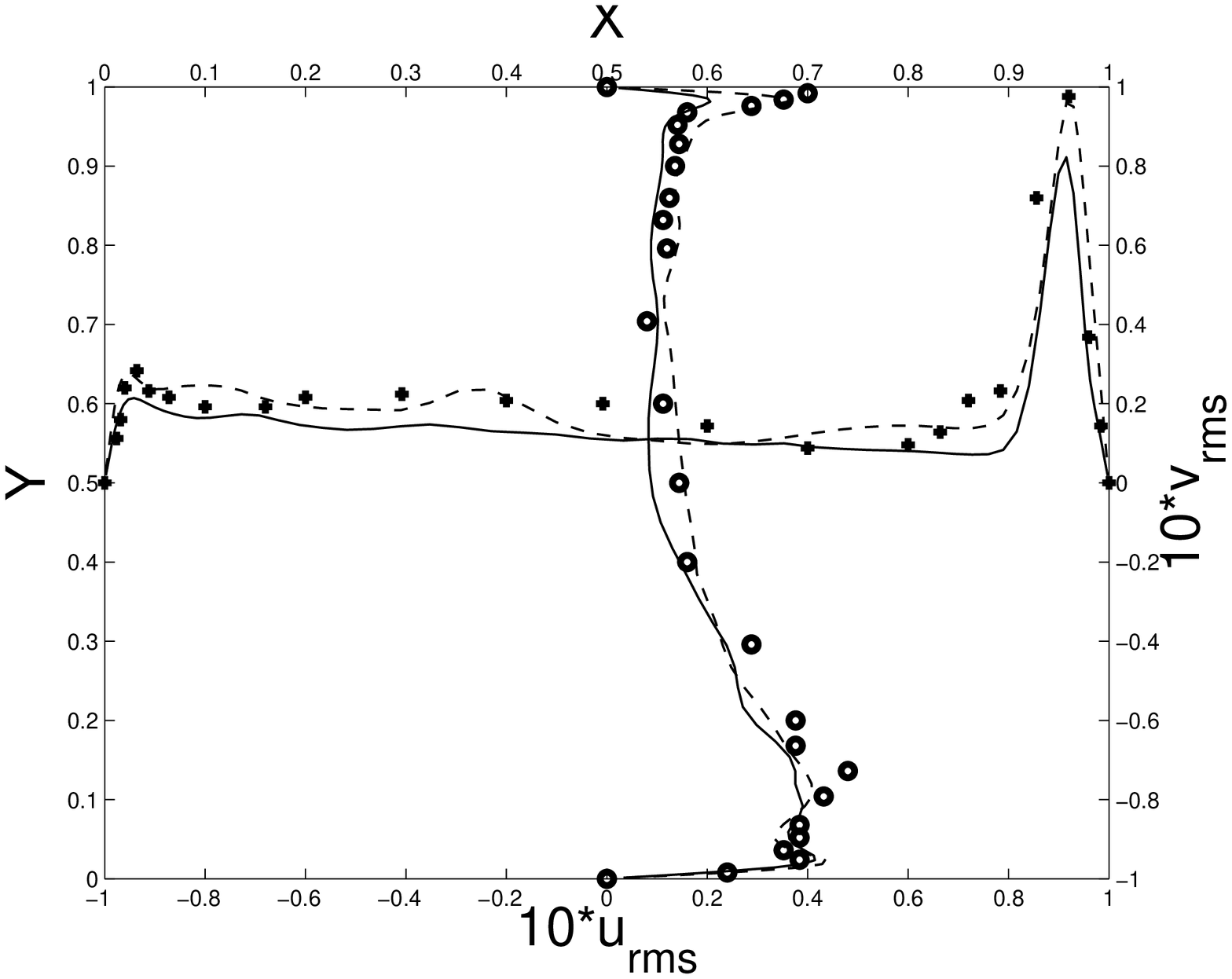,width=7cm}}}
\mbox{
\psfrag{uv}{$500\langle \tl{u}' \tl{v}' \rangle$}
\subfigure[$\langle \tl{u}' \tl{v}' \rangle$ profiles ]{\epsfig{file=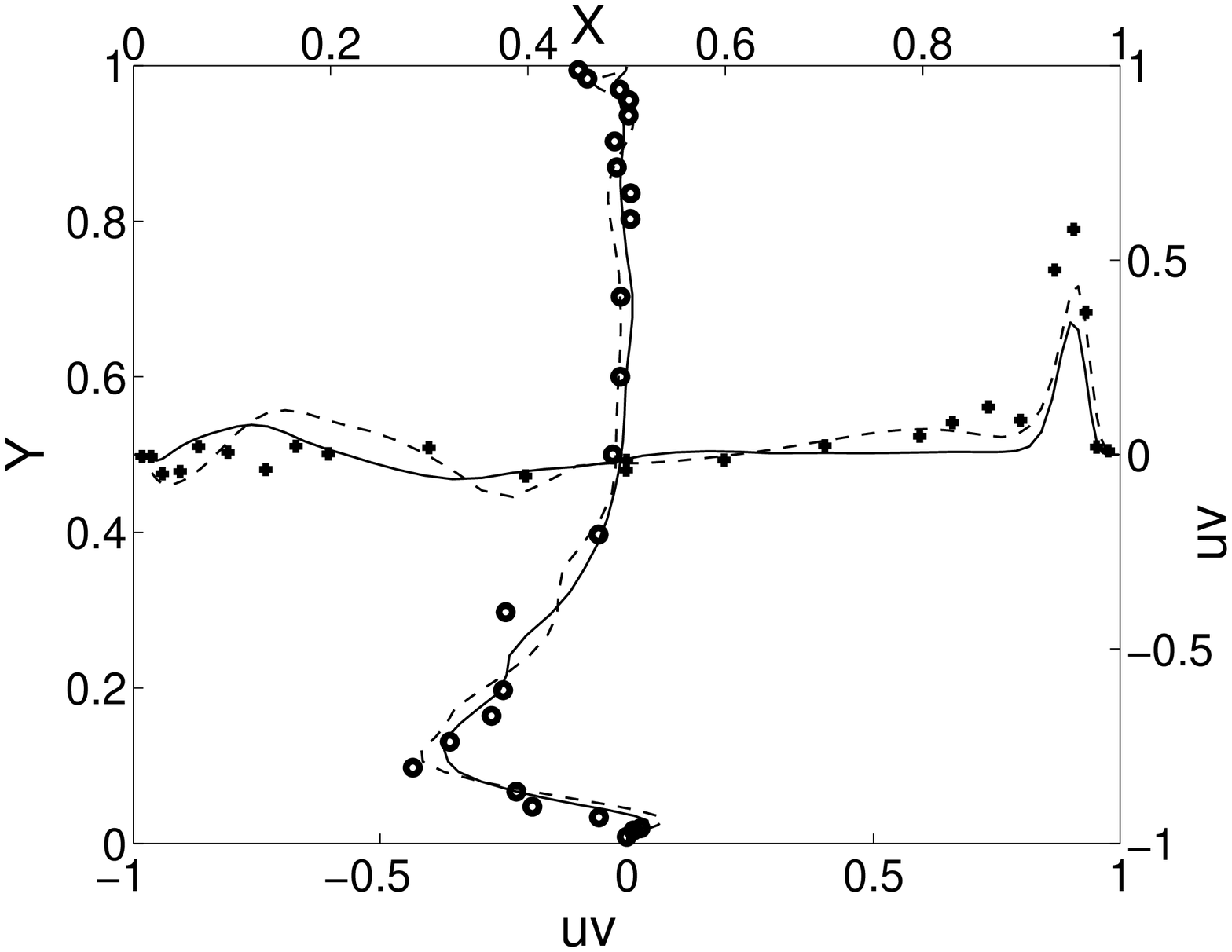,width=7cm}}}
\caption{Mean flow, rms and $\langle \tl{u}'\tl{v}' \rangle$ profiles on the midplane for the $64^{3}$ mesh. Solid line is no model, dashed line is with alternative definition of $\alpha_{k}^{2}$. Symbols are experimental data. \cite{Prasad1989}}
\label{profiles}
\end{figure}

\begin{figure}
\centering
\epsfig{file=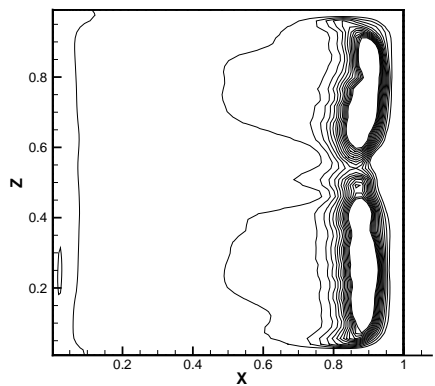,height=6cm}
\caption{$P_{22}$ contours on the $y=0.03$ plane for the $64^{3}$ mesh, levels between $-0.015$ and $0.045$.}
\label{P22}
\end{figure}

\begin{figure}
\centering
\epsfig{file=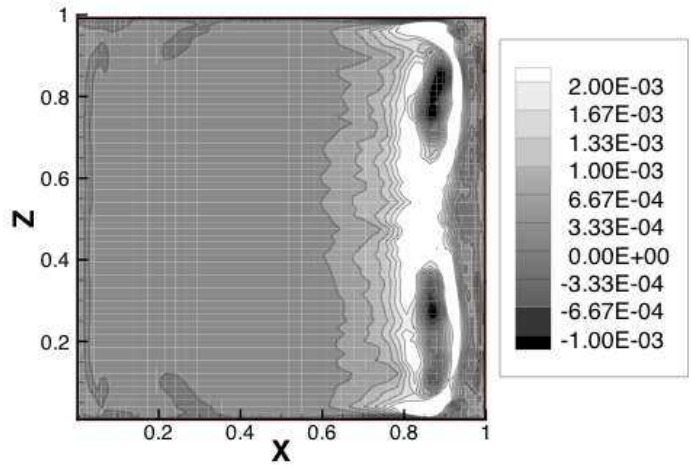,height=7cm}
\caption{Energy transfer term $\tl{u}_{i}\tl{F}_{i}$ on the $y=0.02$ plane for the $64^{3}$ mesh.}
\label{etrans}
\end{figure}

\begin{figure}
\centering
\mbox{
\subfigure[$(\alpha^{2}_{y}/h_{y}^{2})$ on the $z=0.3$ plane ]{\epsfig{file=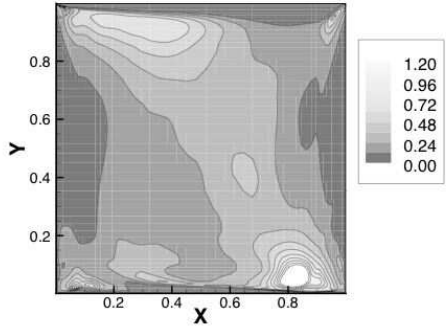,width=7cm}}}
\mbox{
\subfigure[$(\alpha^{2}_{x}/h_{x}^{2})$ on the $y=0.01$ plane ]{\epsfig{file=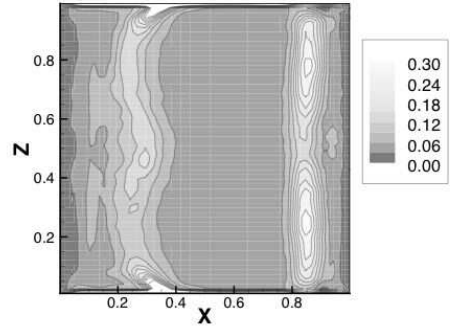,width=7cm}}}
\mbox{
\subfigure[$(\alpha^{2}_{z}/h_{z}^{2})$ on the $y=0.01$ plane ]{\epsfig{file=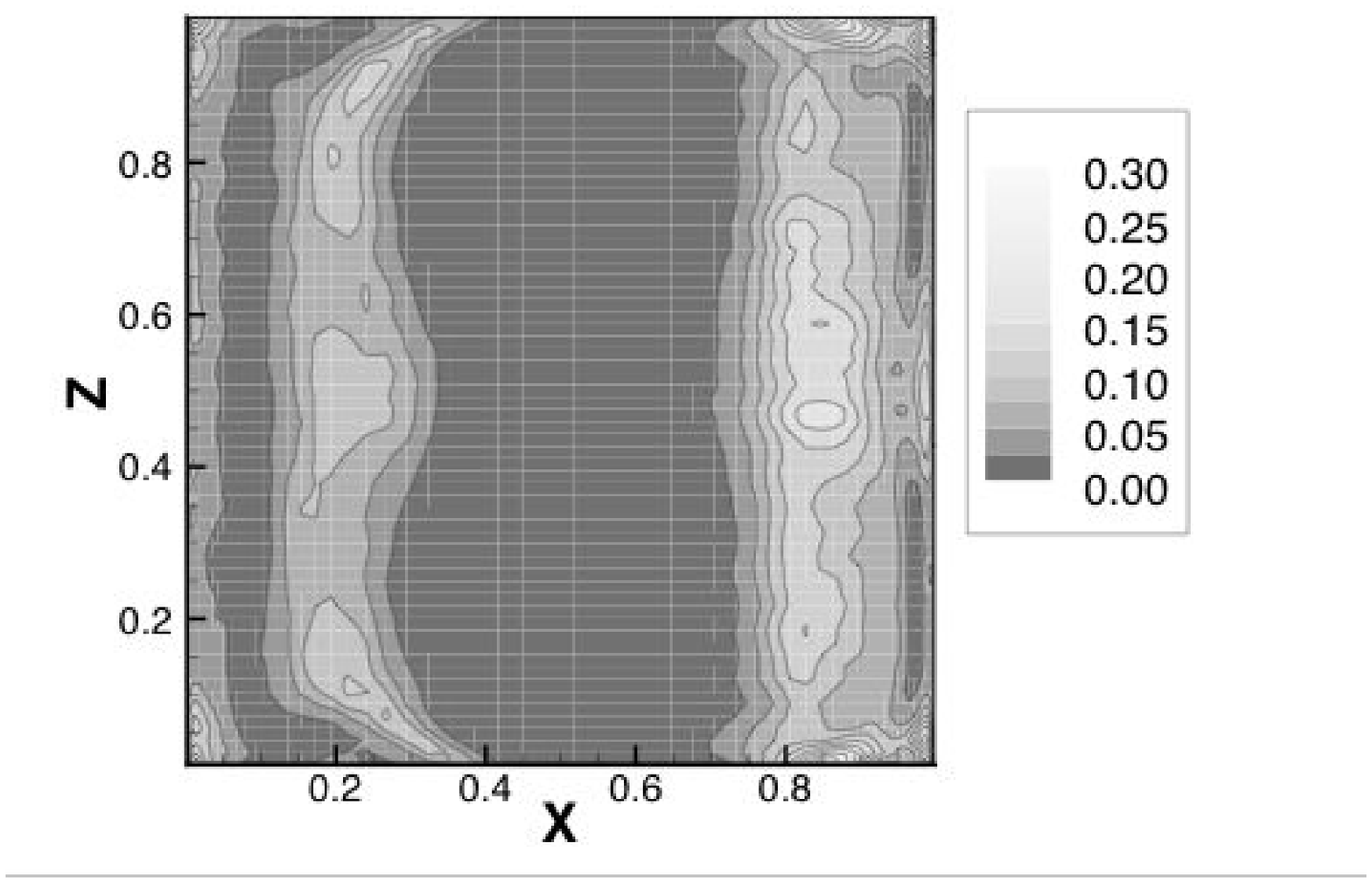,width=7cm}}}
\caption{Contour plots of $\alpha_{j}^{2}/h_{j}^{2}$ highlighting the wall jet impingement and spreading regions for the $64^{3}$ mesh.}
\label{akcontours}
\end{figure}

\begin{figure}
\centering
\mbox{
\subfigure[$\alpha^{2}_{k}$ based on the grid.]{\epsfig{file=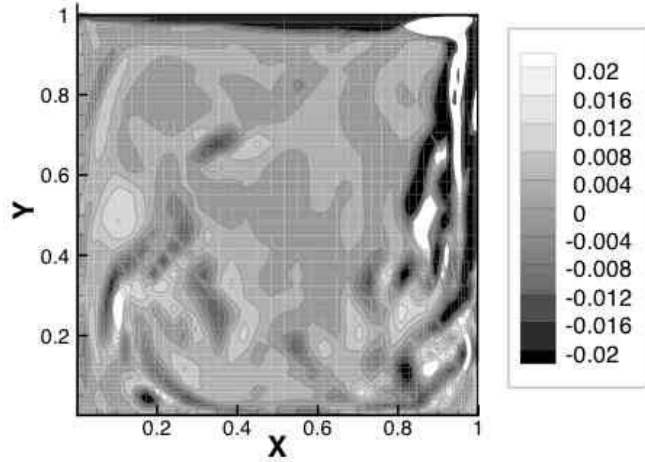,height=7cm}}}
\mbox{
\subfigure[alternative definition of $\alpha^{2}_{k}$ ]{\epsfig{file=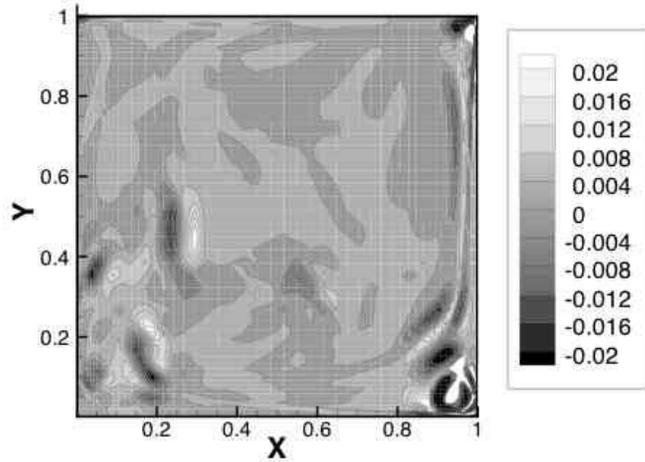,height=7cm}}}
\caption{Subgrid force to the $x-$momentum equation on the $z=0.3$ plane for the $64^{3}$ mesh. With $\alpha^{2}_{k}$ based on the grid the force is high in the laminar regions (near the lid and downstream wall), whereas with the alternative definition (equations \eqref{alphataylor1}-\eqref{alphataylor3}) the force is high only in the turbulent regions.}
\label{sgsforce}
\end{figure}

We now look at the performance of the model with the alternative definition of $\alpha^{2}_{k}$ given in equations \eqref{alphataylor1}-\eqref{alphataylor3}. For comparison, results are also shown for the case where no subgrid model is used. There are several ways the performance of a subgrid model can be assessed. We start by looking at how well the mean flow is captured, which is reflected in the wall jet structure. Recall that the flow should split into two wall jets, which impinge on the cavity bottom. We can see in Figure \ref{walljets} that when a model is not used the flow does not split into two jets, and that this situation is corrected when we used the NS-$\alpha$ model. We found that even on the coarse mesh of $(32)^{3}$ the NS-$\alpha$ model with the alternative definition of $\alpha^{2}_{k}$ can correctly produce the splitting into two wall jets. However, the energy spectra at such a coarse resolution did not exhibit a $k^{-53}$ slope, so no results from this test case are shown.
\\
\\
The mean flow, rms and shear stress profiles are shown in Figure \ref{profiles48} for the $48^{3}$ mesh and in Figure \ref{profiles} for the $64^{3}$ mesh. In Figure \ref{profiles48} we also show the profile from using the mesh-based $\alpha^{2}_{k}$. It can be clearly seen the flow-dependent definition is necessary to obtain the correct mean flow profile. It can also be clearly seen that the new model does a good job of capturing the velocity fluctuations near the lid and in the downstream wall jet region, and that the shear stress profiles are in excellent agreement with the experimental data \cite{Prasad1989}. In contrast, without the model (solid line) the fluctuations are too low, and the shear stress is underpredicted. For the finer mesh results shown in Figure \ref{profiles} the differences with and without the model are small, indicating that as $\alpha^{2}_{k} \rightarrow 0$ the simulation moves towards a DNS as it should. 
\\
\\
The highly inhomogeneous and anisotropic nature of lid driven cavity flow has been was well documented in the DNS and LES studies of Leriche and Gavrilakis \cite{Leriche2000} and Bouffanais and Deville \cite{Bouffanais2007}. One measure of anisotropy they used is the ratios of the volume-averaged contributions of the mean velocity components to the kinetic energy. In the present study it was found the ratio   
$\int_{V} \langle u \rangle ^{2} \,dV : \int_{V} \langle v \rangle ^{2} \,dV : \int_{V} \langle w \rangle^{2} \,dV$ was $1:1.23:118$ without the model as compared to $1:1.21:60$ with the model, both on the $64^{3}$ mesh. The model compares much more favorably with the DNS study which reported $1:1.22:50$. This can be expected from the stronger impingement of the wall jet when the model is used, and the resulting momentum transfer into the spanwise direction. The stronger impingement is very evident if we look at the contours of the production term, $P_{22}=-\langle \tl{v}' \tl{v}' \rangle \partial_{y}\langle \tl{v} \rangle$. The contours shown in Figure \ref{P22} are in good qualitative agreement with the DNS study \cite{Leriche2000}(Figure 14).
\\
\\
Since the flow in the downstream wall jet region is characterized by positive and negative turbulent energy production \cite{Leriche2000} we expect the contribution of the subgrid model to the resolved flow energy equation to exhibit positive and negative values in this region also. The contribution of the subgrid stress to the resolved flow energy equation is
\begin{equation*}
\tl{u}_{i}\frac{\partial m_{ij}}{\partial x_{j}}=\frac{\partial}{\partial x_{j}}\left(\tl{u}_{i}m_{ij}\right)-m_{ij}\frac{\partial \tl{u}_{i}}{\partial x_{j}}.
\end{equation*}
The first term  on the RHS is the transport due to the resolved flow while the second is a source/sink term, usually referred to as the SGS dissipation term. Since it can be both positive or negative, we prefer to call it the SGS transfer term, as it is responsible for the energy transfer between the resolved and subgrid modes (there is an equal and opposite term in the subgrid-scale energy equation \cite{Sagaut2002}). In our method we do not compute $m_{ij}$ explicitly, but rather the volume-integrated subgrid force, 
\begin{equation*}
\tl{F}_{i}=H^{-1}\int_{V}\frac{\partial m_{ij}}{\partial x_{j}}\, dV.
\end{equation*}
This means we cannot split the energy transfer into these two contributions but instead plot the total SGS contribution, $\tl{u}_{i}\tl{F}_{i}$ divided by the control volume. Contour plots of this term on a plane near the cavity bottom are shown in Figure \ref{etrans}. It can be seen there are both negative and positive contributions, and that the impingement points are associated with the energy transfer from the resolved flow, while the spreading is associated with energy transfer to the resolved flow. This is in good agreement with the DNS which found both positive and negative turbulent kinetic energy production terms in this region.
\\ 
\\
To compare the current definition of $\alpha^{2}_{k}$ given in \eqref{alphataylor1}-\eqref{alphataylor3} with the mesh-based definition from equation \eqref{alphamesh}, plots of $\alpha^{2}_{k}/h_{k}^{2}$ are shown in Figure \ref{akcontours}. We can see that $\alpha^{2}_{y}/h^{2}_{y}$ is high in the jet impingement region, while $\alpha^{2}_{x}/h^{2}_{x}$ and $\alpha^{2}_{z}/h^{2}_{z}$ reflect the spreading of the jet on cavity bottom, and the impingement on the upstream wall. Considering that the relationship between the unsmoothed and smoothed velocity in Fourier space is $\hat{u}_{i}(k)=(1+\alpha^{2}k^{2})\hat{\tl{u}}_{i}(k)$ and the maximum resolvable wavenumber is $k \sim \pi/h$ we can also look at this as the range of $\left(\alpha k\right)^{2}$ values. When $\left(\alpha k \right)^{2}=0$ the model is inactive, while in the turbulent regions we expect $\left( \alpha k \right)^{2} \sim 1$. This is reflected in the plots shown in Figure \ref{akcontours}. 
 \\
\\
The actual force experienced by the flow due to the subgrid model is also of interest. In Figure \ref{sgsforce} we plot the subgrid force contribution to the $x-$momentum equation, which can be compared to the mesh based definition discussed earlier. It can be seen that the high source terms near the lid and in the downstream wall jet region are eliminated when the flow dependent version of $\alpha_{k}^{2}$ is used, and instead the flow is active in the turbulent regions near the cavity bottom. 

\section{Conclusions }
\label{conclusions}
An anisotropic version of the NS-$\alpha$ subgrid model (where $\tl{u}$ is the dependent variable) was developed starting from the anisotropic Eulerian-averaged equations given by Holm \cite{Holm1999} in a manner that should be familiar to the LES community. While simplifications were made (Section 2), this work here still represents (to the author's knowledge) the first application of the anisotropic NS-$\alpha$ equations as a subgrid model in the context of LES, including the solution of a wall-bounded flow.  Because the isotropic equations are showing promise in unbounded flows, we view this as a first step towards a more general application of the model to complex flows. The full anisotropic subgrid model should be tested against the one used here (with only the diagonal $\alpha^{2}_{k}$) using a more appropriate test case in a separate study.
\\
\\
The model was found to be sensitive to abrupt changes in $\alpha^{2}$. This is not surprising since $\alpha^{2}$ is supposed to be a smoothing parameter, and abrupt changes are hardly physical. However, if $\alpha^{2}=0$ on the solid boundary is to be enforced, it was found this can be a problem. For the three-dimensional cavity flow this problem manifested itself in the form of oscillations in the spanwise velocity field and in the appearance of small-scale vertical vorticity. This vorticity can be understood as being due to the tilting of the spanwise vorticity from the Stokes-vortex force, an effect here which is numerical rather than physical. To overcome this problem an alternative definition of $\alpha^{2}_{k}$ was proposed which is not based solely on the mesh spacing. This definition worked very well in capturing the wall-jet splitting, rms and shear stress profiles, and was also found to predict both forward energy transfer and backscatter in the jet impingement regions in qualitative agreement with the discussion given in Leriche and Gavilakis \cite{Leriche2000}. The alternative definition allows us to use the model in a complex flow situation that presents a significant challenge to most subgrid models. For the lid-driven cavity it was important that the model remain inactive in the laminar flow regions, which was not possible when $\alpha^{2}$ was based on the mesh. While this was not a problem in the mixing layer study carried out by Geurts and Holm \cite{Geurts2006}, it should be noted their problem did not have solid boundaries, and was relatively symmetric in the early stages of transition. 
\\
\\
Lastly it should be mentioned that simulations were also done with the $C_{ij}$ term turned off, which is similar to using a Leray model. It was found in these cases that there was no benefit to using the model, and in some cases the model tended to damp the small scale activity strongly. This is in agreement with recent results \cite{Graham2008} which indicate the Leray model reduces the effective Reynolds number of the flow. The tilting term, $u_{k}\partial_{i}\tl{u}_{k}$, which combines with the modified pressure gradient to form the $\partial_{j}(C_{ij})$ term in the model, is the unique feature of the NS-$\alpha$ model. The role of this term is presently being investigated in turbulent channel flows. It is hoped the channel flow cases will also delineate the near-wall behavior of the model further.

\clearpage
\begin{table}[b]
\begin{center}
\begin{tabular}{|l|c|c|c|c|c|}
\hline
 Re       &  (Nx,Ny,Nz) & $\Delta_{min}$ & $\Delta_{max}/\Delta_{min}$ \\
\hline
 10, 000 &   (32,32,32) & $5.3 \times10^{-3}$ & 15.7\\
 \hline 
 10, 000 & (48,48,48) & $3.6 \times10^{-3}$ & 12.5 \\
\hline
 10, 000 & (64,64,64)  &  $2.6 \times 10^{-3}$ & 12.9 \\
\hline
\end{tabular}
\caption{Mesh parameters}
\end{center}
\end{table}
\vspace{5cm}

%
%
%

\newpage

%
%
%

%
%
%

\clearpage


\bibliographystyle{plain}       
\bibliography{cavitye.bib}   

\begin{thebibliography}{10}

\bibitem{Batchelor1972}
G.K. Batchelor.
\newblock {\em {The Theory of Homogeneous Turbulence}}.
\newblock {Cambridge University Press}, 1972.

\bibitem{Bouffanais2007}
R.~Bouffanais and M.~O. Deville.
\newblock Large-eddy simulation of the flow in a lid-driven cavity.
\newblock {\em Physics of Fluids}, 19:055108, 2007.

\bibitem{Chen1998}
S.~Chen, C.~Foias, D.D. Holm, E~Olson, E.S. Titi, and S.~Wynne.
\newblock Camassa-{H}olm equations as a closure model for turbulent channel and
  pipe flow.
\newblock {\em Physical Review Letters}, 81:5338--5341, 1998.

\bibitem{Chen1999a}
S.~Chen, C.~Foias, D.D. Holm, E.~Olson, E.S. Titi, and S.~Wynne.
\newblock The {C}amassa-{H}olm equations and turbulence.
\newblock {\em Physica D}, 133:49--65, 1999.

\bibitem{Chen1999c}
S.~Chen, C.~Foias, D.D. Holm, E.~Olson, E.S. Titi, and S.~Wynne.
\newblock A connection between the {C}amassa-{H}olm equations and turbulent
  flows in pipes and channels.
\newblock {\em Physics of Fluids}, 11:2343--2353, 1999.

\bibitem{Chen1999b}
S.~Chen, D.D. Holm, L.G. Margolin, and R.~Zhang.
\newblock {D}irect numerical simulations of the {N}avier-{S}tokes alpha model.
\newblock {\em Physica D}, 133:66--83, 1999.

\bibitem{Craik1976}
A.D.D. Craik and S.~Leibovich.
\newblock A rational model for {L}angmuir circulation.
\newblock {\em Journal of Fluid Mechanics}, 73:401--426, 1976.

\bibitem{Domaradzki2001}
J.A. Domaradzki and D.D. Holm.
\newblock Navier-{S}tokes alpha model: {L}{E}{S} equations with nonlinear
  dispersion.
\newblock In B.J. Geurts, editor, {\em Modern Simulation Strategies for
  Turbulent Flow}, chapter~6. R.T. Edwards, Inc., 2001.

\bibitem{Ducros1996}
{F. Ducros P. Comte and M. Lesieur}.
\newblock Large-eddy simulation of transition to turbulence in a boundary layer
  developing spatially over a flat plate.
\newblock {\em Journal of Fluid Mechanics}, 326:1--36, 1996.

\bibitem{Foias2001}
C.~Foias, D.D. Holm, and E.S. Titi.
\newblock The {N}avier-{S}tokes-alpha model of fluid turbulence.
\newblock {\em Physica D}, 152-153:505--519, 2001.

\bibitem{Freitas1988}
C.J. Freitas and R.L. Street.
\newblock Non-linear transient phenomena in a complex recirculating flow: {A}
  numerical investigation.
\newblock {\em International Journal for Numerical Methods in Fluids},
  8:769--802, 1988.

\bibitem{Geurts2003}
B.J. Geurts.
\newblock {\em Elements of Direct and Large-Eddy Simulation}.
\newblock R.T.Edwards, 2003.

\bibitem{Geurts2006}
B.J. Geurts and D.D. Holm.
\newblock Leray and {LANS}-alpha modelling of turbulent mixing.
\newblock {\em Journal of Turbulence}, 7(10):1--33, 2006.

\bibitem{Graham2008}
J.~Graham, D.~Holm, P.~Mininni, and A.~Pouquet.
\newblock Three regularization models of the {N}avier-{S}tokes equations.
\newblock {\em Physics of Fluids}, 20:035107, 2008.

\bibitem{Hanjalic2005}
K.~Hanjalic.
\newblock {Will RANS survive LES: a view of perspectives}.
\newblock {\em Journal of Fluids Engineering}, 127:831--839, 2005.

\bibitem{Holm1999}
D.D. Holm.
\newblock Fluctuation effects on {3D} {L}agrangian mean and {E}ulerian mean
  fluid motion.
\newblock {\em Physica D}, 133:215--269, 1999.

\bibitem{Holm2005}
D.D. Holm.
\newblock Taylor's hypothesis, {H}amilton's principle and the
  {L}{A}{N}{S}-alpha model for computing turbulence.
\newblock {\em Los Alamos Science}, (29):172--180, 2005.

\bibitem{Holm2003}
D.D. Holm and B.T. Nadiga.
\newblock Modeling mesoscale turbulence in the barotropic double-gyre
  circulation.
\newblock {\em Journal of Physical Oceanography}, 33:2355--2366, 2003.

\bibitem{Leibovich1983}
S.~Leibovich.
\newblock The form and dynamics of {L}angmuir circulations.
\newblock {\em Annual Review of Fluid Mechanics}, 15:715--724, 1983.

\bibitem{Leray1934}
J.~Leray.
\newblock Sur les movements d'un fluide visqueux remplissant l'espace.
\newblock {\em Acta Mathematica}, 63:193--248, 1934.

\bibitem{Leriche2000}
E.~Leriche and S.~Gavrilakis.
\newblock Direct numerical simulations of the flow in a lid-driven cubical
  cavity.
\newblock {\em Physics of Fluids}, 12:1363, 2000.

\bibitem{Lesieur1996}
M.~Lesieur and O.~Metais.
\newblock New trends in large-eddy simulations of turbulence.
\newblock {\em Annual Review of Fluid Mechanics}, 28:45--82, 1996.

\bibitem{Lien1994a}
F.S. Lien and M.A. Leschziner.
\newblock A general non-orthogonal collocated {F}{V} algorithm for turbulent
  flow at all speeds incorporating second moment closure. {P}art 1:
  {C}omputational implementation.
\newblock {\em Computer Methods for Applied Mechanics and Engineering},
  114:123--148, 1994.

\bibitem{Marsden2003}
J.E. Marsden and S.~Shkoller.
\newblock The {A}nisotropic {L}agrangian {A}veraged {E}uler and
  {N}avier-{S}tokes equations.
\newblock {\em Archives of Rational Mech. Analysis}, 66:27--46, 2003.

\bibitem{McWilliams1997}
J.C. McWilliams, P.P. Sullivan, and C.H. Moeng.
\newblock Langmuir turbulence in the ocean.
\newblock {\em Journal of Fluid Mechanics}, 334:1--30, 1997.

\bibitem{Mohseni2003}
K.~Mohseni, B.~Kosovic, S.~Shkoller, and J.E. Marsden.
\newblock Numerical simulations of the {L}agrangian {A}veraged
  {N}avier-{S}tokes equations for homogeneous isotropic turbulence.
\newblock {\em Physics of Fluids}, 15(2):524--544, 2003.

\bibitem{Petersen2008}
M.R. Petersen, M.W. Hecht, and B.A. Wingate.
\newblock Efficient form of the {LANS}-alpha turbulence model in a primitive
  equation ocean model.
\newblock {\em Journal of Computational Physics (accepted)}, 2008.

\bibitem{Piomelli1991}
U.~Piomelli, W.H. Cabot, P.~Moin, and S.~Lee.
\newblock Subgrid-scale backscatter in turbulent and transitional flows.
\newblock {\em Physics of Fluids A}, 7(3):1766--1771, 2001.

\bibitem{Prasad1989}
A.K. Prasad and J.R. Koseff.
\newblock Reynolds number and end-wall effects on a lid-driven cavity flow.
\newblock {\em Physics of Fluids A}, 1(2):208--218, 1988.

\bibitem{Sagaut2002}
P.~Sagaut.
\newblock {\em Large Eddy Simulation for Incompressible Flows}.
\newblock Springer-Verlag, 2002.

\bibitem{Taylor1922}
G.I. Taylor.
\newblock Diffusion by continuous movements.
\newblock {\em Proceedings of the London Mathematical Society}, 20:196--212,
  1922.

\bibitem{vanderBos2005}
F.~van~der Bos and B.J. Geurts.
\newblock Commutator errors in the filtering approach to large-eddy simulation.
\newblock {\em Physics of Fluids}, 17:035108, 2005.

\bibitem{vanReeuwijk2007}
M.~van Reeuwijk.
\newblock {\em Direction simulation and regularization modeling of turbulent
  thermal convection}.
\newblock PhD thesis, Delft University of Technology, 2007.

\bibitem{vanReeuwijk2006}
M.~van Reeuwijk, H.J.J. Jonker, and K.~Hanjalic.
\newblock Incompressibility of the {L}eray-alpha model for wall-bounded flows.
\newblock {\em Physics of Fluids}, 18:018103, 2006.

\bibitem{Verstappen2007}
Verstappen2007.
\newblock Symmetry-preserving regularization modeling of turbulent channel
  flow.
\newblock In {\em Turbulent Boundary Layers}, June 2007.

\bibitem{Vreman2004}
B.~Vreman.
\newblock Comment on '{I}napplicability of the dynamic {C}lark model to the
  large eddy simulation of incompressible turbulent channel flows.
\newblock {\em Physics of Fluids}, 16(2):L29, 2004.

\bibitem{Vreman1996}
B.~Vreman, B.J. Geurts, and H.~Kuerten.
\newblock Large-eddy simulation of the temporal mixing layer using the {C}lark
  model.
\newblock {\em Theoretical and Computational Fluid Dynamics}, 8:309--324, 1996.

\bibitem{Winckelmans2001}
G.S. Winckelmans, O.~Wray, A.A.~Vasilyev, and H.~Jeanmart.
\newblock Explicit-filtering large-eddy simulation using the tensor-diffusivity
  model supplemented by a dynamic {S}magorinsky term.
\newblock {\em Physics of Fluids}, 13(5):1385--1403, 2001.

\bibitem{Zang1993}
Y.~Zang, R.L. Street, and J.R. Koseff.
\newblock A dynamic mixed subgrid-scale model and its application to turbulent
  recirculating flows.
\newblock {\em Physics of Fluids A}, 5(12):3186--3196, 1993.

\bibitem{Zhao2005b}
H.~Zhao and K.~Mohseni.
\newblock Anisotropic turbulent flow simulations using the
  {L}agrangian-{A}veraged {N}avier-{S}tokes alpha equation.
\newblock In {\em Proceedings of the 15th AIAA Fluid Dynamics conference and
  Exhibit}, June 2005.

\bibitem{Zhao2005a}
H.~Zhao and K.~Mohseni.
\newblock A dynamic model for the {L}agrangian {A}veraged {N}avier-{S}tokes
  $\alpha$ equations.
\newblock {\em Physics of Fluids}, 17:075106, 2005.

\end{thebibliography}


\end{document}